\pdfoutput=1
\documentclass[final,3p,times]{elsarticle}
\usepackage{graphicx}
\usepackage{amssymb}
\usepackage{bm}

\newcommand{\Uad}{U_\mathrm{ad}}
\newcommand{\UCP}{U_\mathrm{CP}}

\newcommand{\Rb}{^{87}\mathrm{Rb}}
\newcommand{\be}{\begin{equation}}
\newcommand{\ee}{\end{equation}}

\journal{Comptes Rendus}

\begin{document}

\begin{frontmatter}

\title{Coupling ultracold atoms to mechanical oscillators}

\author[label1,label2]{D. Hunger}
\author[label1,label2]{S. Camerer}
\author[label1,label2,label3]{M. Korppi}
\author[label1,label2,label3]{A. J{\"o}ckel}
\author[label1,label2]{T.~W. H{\"a}nsch}
\author[label1,label2,label3]{P. Treutlein}

\address[label1]{Ludwig-Maximilians-Universit{\"a}t M{\"u}nchen, Schellingstr. 4, 80799 M{\"u}nchen, Germany}
\address[label2]{Max-Planck-Institut f{\"u}r Quantenoptik, Hans-Kopfermann-Str. 1, 85748 Garching, Germany}
\address[label3]{Departement Physik, Universit{\"a}t Basel, Klingelbergstrasse 82, 4056 Basel, Switzerland}

\begin{abstract}
In this article we discuss and compare different ways to engineer an interface between ultracold atoms and micro- and nanomechanical oscillators. We start by analyzing a direct mechanical coupling of a single atom or ion to a mechanical oscillator and show that the very different masses of the two systems place a limit on the achievable coupling constant in this scheme. We then discuss several promising strategies for enhancing the coupling: collective enhancement by using a large number of atoms in an optical lattice in free space, coupling schemes based on high-finesse optical cavities, and coupling to atomic internal states. Throughout the manuscript we discuss both theoretical proposals and first experimental implementations.
\end{abstract}


\end{frontmatter}

\section{Introduction}\label{sec:Intro}

Recent experiments have demonstrated an impressive level of control over micro- and nanomechanical oscillators.
One key achievement is the engineering of mechanical modes with extremely low dissipation, providing remarkable isolation from the environment \cite{Verbridge08,Zwickl08,Wilson09}. 
A second cornerstone is the detection and manipulation of micro- and nanomechanical motion. Both in solid-state based systems \cite{Schwab05} and cavity optomechanical settings \cite{Kippenberg08,Marquardt09,Favero09b}, techniques have been perfected to monitor position fluctuations with an imprecision down to the standard quantum limit \cite{Arcizet06,Abbott09,Rocheleau09,Hertzberg09,Schliesser09,Groeblacher09b,Teufel09}. 
Moreover, sophisticated techniques for cooling and coherent excitation of individual mechanical modes have been developed \cite{Arcizet06,Abbott09,Schliesser09,Groeblacher09b,HoehbergerMetzger04,Teufel08}, providing access to the quantum ground state \cite{OConnell10}.

In the light of these achievements, mechanical oscillators appear as promising systems for the study of quantum physics. 
The motivation is manifold. Firstly, manipulations on the quantum level represent the ultimate control over a system. Practically, this would have implications for the achievable sensitivity e.g.\ in force sensing \cite{Ekinci05}, gravitational wave sensing \cite{Caves81}, or nuclear spin imaging \cite{Degen09}.
Secondly, quantum control over massive objects could enable new tests of non-standard decoherence theories \cite{Ghirardi86,Penrose00} and illucidate the crossover from microscopic quantum systems to macroscopic classical systems.
Finally, mechanical oscillators coherently coupled to other well-controlled quantum systems are examples of \textit{hybrid quantum systems}, which are of interest in the context of quantum engineering, quantum information networking, and quantum information processing \cite{Rabl10}.

The different systems that are investigated have specific advantages and drawbacks.
Optomechanical systems so far mostly realize a linear coupling of mechanical motion to a coherently excited optical cavity mode. In this limit, the coupling can be very strong, and e.g.\ a mechanical analog of the Autler-Townes splitting in the energy spectrum of a coupled resonator-cavity system was observed \cite{Groeblacher09} (similarly in an electromechanical system \cite{Teufel10}). Theoretical proposals have shown that the coupled system could generate mechanical squeezing \cite{Fabre94,Jaehne09,Chang10}, entanglement between the light field and the resonator \cite{Chang10,Pinard05,Vitali07}, or Schr{\"o}dinger cat states of the light field \cite{Mancini97}.
However, the creation of mechanical Fock states or more general quantum superpositions is not possible by a linear coupling to a classical field. Different approaches have been developed to overcome this limitation. For example, dispersive measurements that rely on a quadratic coupling to the light field could permit the observation of quantum jumps between Fock states \cite{Thompson08,Jayich08}. Alternatively, non-classical light states such as single photon states could be employed to generate superpositions of position states \cite{Bose99,Marshall03}. Yet, single photons lead only to a minute coupling, and enhancement of the interaction by additional classical fields is required to achieve strong coupling with current experimental parameters \cite{Groeblacher09,RomeroIsart10,RomeroIsart10b,Akram10}.

A promising way to achieve full quantum control of a mechanical oscillator is to couple it to another well-controlled quantum system, preferably a two-level system.
Solid-state two-level systems are very attractive in this context, and a variety of scenarios has been discussed \cite{Armour02,Bargatin03,Cleland04,Rugar04,Tian05,Rabl09}. 
Mechanical systems have already been successfully coupled to a single electron spin in a solid \cite{Rugar04} or to a cooper pair box \cite{LaHaye09}, and the recent demonstration of quantum control of a mechanical oscillator coupled to a superconducting qubit \cite{OConnell10} has set the pace in the field. The latter experiment displays the advantages of entirely solid-state based approaches, being in essence a high coupling rate and a natural compatibility with cryogenic environments.
Yet, the coherence times of solid-state based two-level systems as used so far reside in the nanosecond to microsecond regime, which leads to the requirement of an exceptionally high coupling rate in the MHz range and consequently also very high mechanical resonance frequencies.
Furthermore, the mechanical oscillator is typically integrated into a complex electronic environment and carries electrodes, making it difficult to independently optimize the mechanical quality factor.

Coupling ultracold atoms to mechanical oscillators creates a novel, qualitatively different setting. Both optically or magnetically trapped clouds of ultracold atoms as well as trapped ions represent systems where quantum control has been achieved over all degrees of freedom \cite{Chu02}. One of their outstanding features is the possibility to achieve exceptionally long coherence times. Superpositions of internal states have been demonstrated to preserve coherence for several seconds \cite{Harber02,Treutlein04,Langer05,Deutsch10}. Remarkably, for neutral atoms this is still true in close proximity to chip surfaces \cite{Treutlein04}.
In addition, an elaborate toolbox for the quantum manipulation of the atomic state is at hand \cite{Chu02,Bloch05,Hofferberth06,Boehi09}, offering control over internal and motional degrees of freedom.
Atoms can be regarded as mechanical oscillators themselves. Acting as a dispersive medium inside an optical cavity, an interesting variant of dispersive cavity optomechanics can be realized \cite{Gupta07,Brennecke08,Murch08,Purdy10,Brahms10,Kanamoto10}.
Atoms provide both the continuous degree of freedom of collective atomic motion as well as a discrete set of internal levels that can be reduced to a two-level system.
Furthermore, a unique feature of ultracold atoms is that dissipation can be tuned, e.g.\ by switching on/off laser cooling \cite{Chu02}, and that self-interactions can be tuned and used for the generation of entanglement \cite{Gross10,Riedel10}.
The challenge in the present context is to find a suitable coupling mechanism that provides controlled and sufficiently strong interactions of the atoms with a single mechanical mode of a micro- or nanostructured oscillator. 

\subsection*{Coupling mechanisms}
In this article we discuss different approaches to engineer a coupling between atomic systems and micro- and nanomechanical oscillators.

We focus initially on the most straightforward approach, the direct mechanical coupling of the oscillator's motion to the motion of trapped atoms. The coupling results from a force that depends on the distance between the two systems, such that the force gradient converts an oscillation of the position of one system into a modulation of the force experienced by the other.
Energy can be coherently exchanged when both systems oscillate with about the same frequency, i.e. for (near-)resonance conditions.

To realize a direct mechanical coupling, it has been suggested to use electrostatic forces to couple to trapped ions \cite{Wineland98,Hensinger05} or polar molecules \cite{Singh08,Bhattacharya10}. With neutral atoms, a first experiment was already performed \cite{Hunger10a}, exploiting attractive surface forces between a micromechanical oscillator and a Bose-Einstein condensate (BEC), see Sec.~\ref{sec:SurfaceCoupling}. A beneficial variant of this approach is to use the mechanical oscillator as an integral part of the trap itself, either as an electrode of an ion trap \cite{Tian04}, current carrying wire of a magnetic trap \cite{Peano05}, or as a mirror generating a one-dimensional optical lattice \cite{Hammerer10}. We discuss the direct mechanical coupling in detail in Sec.~\ref{sec:DirectCoupling}, where we show that the large mass difference between a single atom or ion and a micro- or nanostructured mechanical oscillator results in an impedance mismatch that limits the coupling to rather small values. We point out that the coupling cannot be simply increased beyond a certain level by increasing the strength of the coupling force, because a significant distortion of the atom trap would result.
However, one can think of several promising strategies to enhance the coupling, and we present them in the context of specific implementations in the subsequent chapters.

We first discuss the possibility to compensate the trap deformation in order to achieve stronger coupling in the case of an ion trap (Sec.~\ref{sec:IonCoupling}). Second, a way to minimize the impedance mismatch is to use molecular-scale oscillators with low mass. We discuss this perspective for surface-force coupling between a BEC and a carbon nanotube (Sec.~\ref{sec:CNTCoupling}). Alternatively, one can increase the effective mass of the atoms by using a large number of them. Such collective enhancement is a promising route to achieve strong coupling mediated by an optical lattice (Sec.~\ref{sec:LatticeCoupling}). 

For the perspective of creating a coupled quantum system where the mechanical oscillator is ``macroscopic'', it is highly desirable to find coupling mechanisms where the impedance mismatch does not play a role. Indeed this is possible in several schemes:
A powerful method is to use a high-finesse optical cavity that incorporates both the mechanical oscillator and the atoms, such that the two systems are coupled via the intracavity light field.
Possible schemes rely either on a dispersive optomechanical coupling \cite{Meiser06,Bhattacherjee09,Hammerer09b,Wallquist10,Zhang10} or on resonant interaction \cite{Genes08,Ian08,Chang09}.
We discuss one particular scenario of this kind \cite{Hammerer09b} (Sec.~\ref{sec:CavityCoupling}), where strong coupling may be achievable even for a single atom and a rather massive membrane oscillator.

Coupling the oscillator motion to the internal degrees of freedom of atoms is another way to avoid the impedance mismatch. This can be achieved by a magnet on the oscillator tip which transduces the mechanical motion into magnetic field oscillations, thereby establishing a coupling to the atomic magnetic moment \cite{Treutlein07}. We discuss this scheme in Sec.~\ref{sec:MagneticCoupling}. It could be used for quantum state tomography of the mechanical oscillator via an atom \cite{Singh10}, enable the generation of entanglement between two nanomechanical oscillators \cite{Joshi10}, and it could be scaled up by the use of atoms in an optical lattice coupling to an array of oscillators \cite{Geraci09}. Magnetic coupling was the first mechanism that was studied experimentally, using an atomic gas in a vapor cell \cite{Wang06}. We note that magnetic coupling is also considered for the coupling of mechanical oscillators to color centers in diamond \cite{Rabl09}.
Finally, the mechanical oscillator could also act on the polarization state of light, which then couples to internal states of atoms \cite{Hammerer09}. The remarkable feature of this scheme is that long-distance entanglement between the mechanical oscillator and the atoms can be created by means of a projective measurement even for a thermal oscillator.

\section{Direct mechanical coupling}\label{sec:DirectCoupling}
\begin{figure}[t]
\centering
\includegraphics[width=0.49\textwidth]{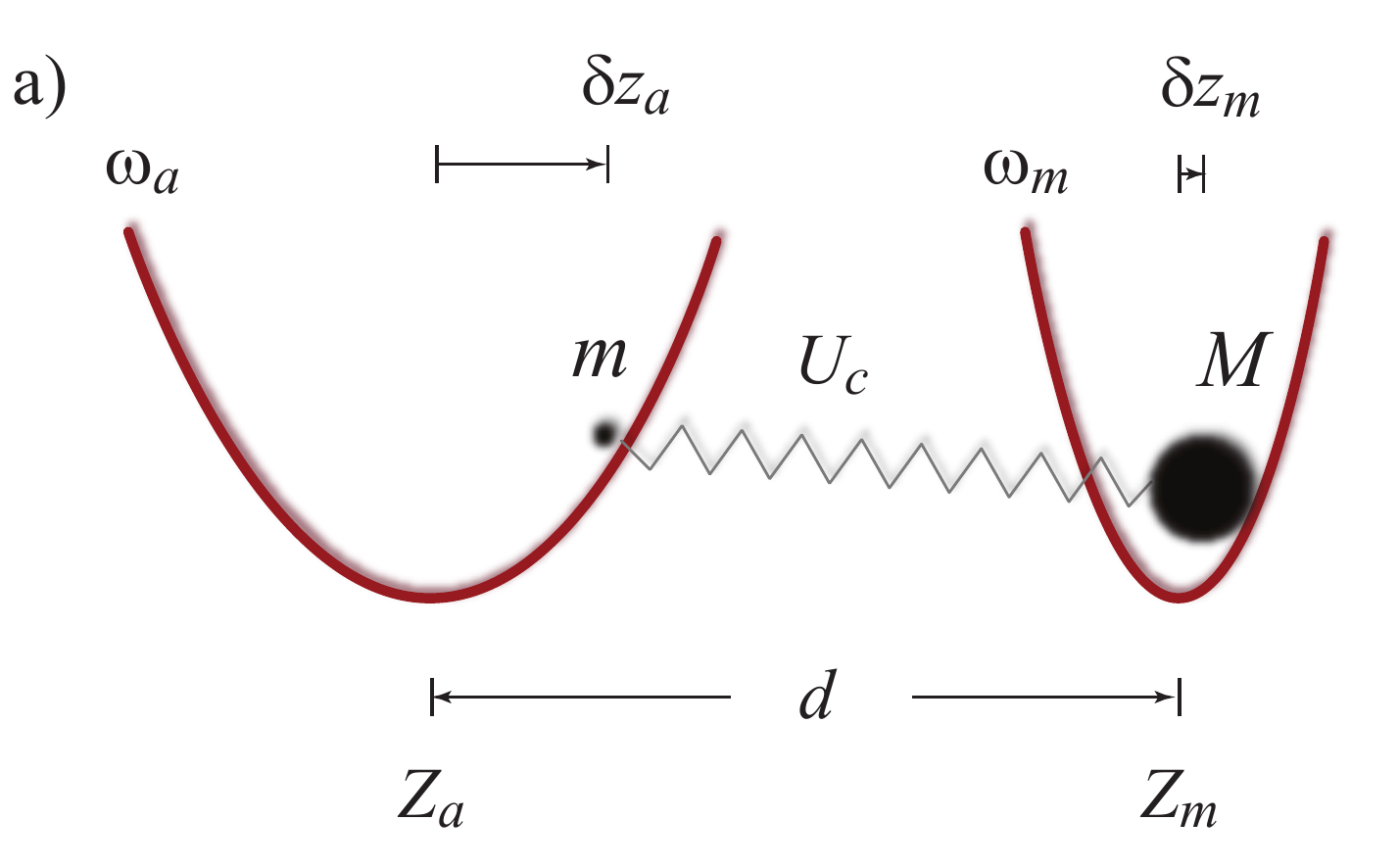}
\includegraphics[width=0.33\textwidth]{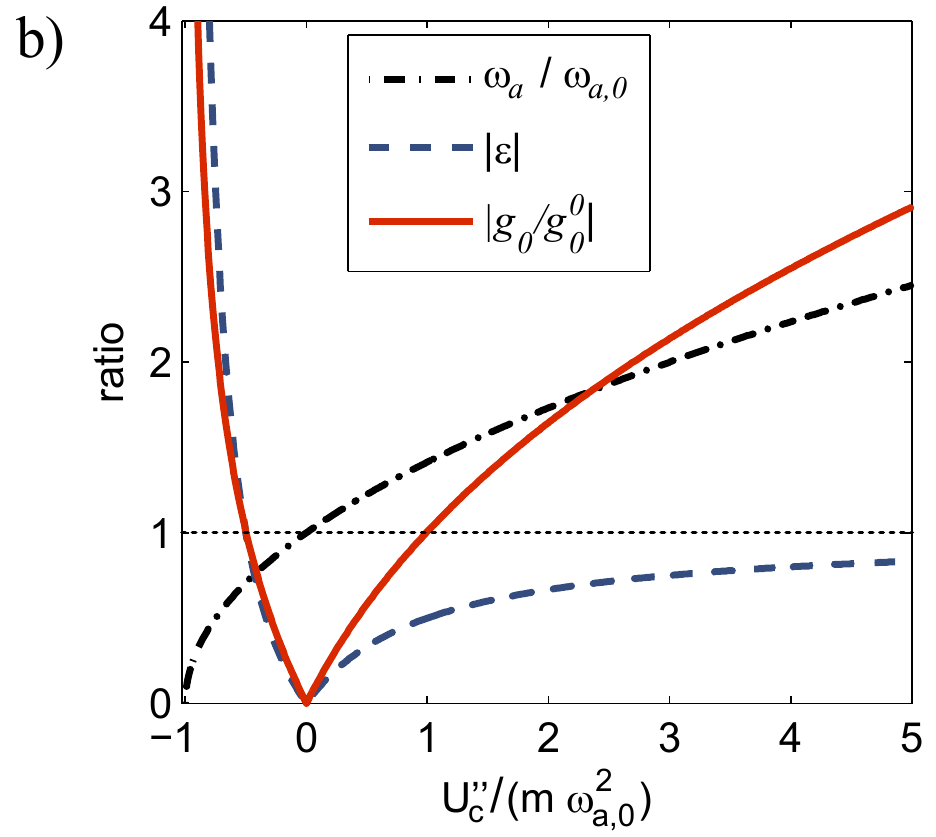}
\caption{\label{fig:CouplingPot}
(a) Direct mechanical coupling of two harmonic oscillators of different mass. The two oscillators with respective masses $m$ and $M$ interact via a potential $U_c$ that depends on their relative distance. The interaction leads to shifted eigenfrequencies $\omega_a$ and $\omega_m$ and an equilibrium distance $d$ (see Sec.~\ref{sec:TrapDeform}). (b) Coupling parameters as a function of the normalized curvature of the coupling potential. We show the relative change in trap frequency $\omega_a/\omega_{a,0}$, the coupling strength parameter $\epsilon$ which compares the curvature of the coupling potential to the curvature of the trap, and the coupling strength $g_0$ normalized to the value $g_0^0$ where $U_c''/m\omega_a^2=1$ (see text).
}
\end{figure}

Let us consider a system of two harmonic oscillators of different mass that are coupled by a spring. 
This is a general model for various proposed methods to directly couple the motion of trapped atoms or ions to the vibrations of a mechanical oscillator via a distance-dependent force. It allows us to make general statements about the achievable coupling strength. 
For simplicity, we initially restrict the discussion to two undamped harmonic oscillators, one being an atom of mass $m$ trapped in a conservative potential with trap frequency $\omega_{a,0}$, and the other being a mechanical oscillator with effective mass $M\gg m$ and eigenfrequency $\omega_{m,0}$, see Fig.~\ref{fig:CouplingPot} (a).
Furthermore, the two oscillators should have similar frequencies $\omega_{a,0}\approx\omega_{m,0}$ to achieve resonant coupling.

The Hamiltonian of the coupled system is given by
\begin{equation}
H =  \frac{p_m^2}{2M} + \frac{1}{2}M\omega_{m,0}^2 (z_m-Z_{m,0})^2 + \frac{p_a^2}{2m} + \frac{1}{2}m\omega_{a,0}^2 (z_a-Z_{a,0})^2 + U_c[z_m-z_a],
\end{equation}
where $p_a$ ($p_m$) is the momentum of the atom (mechanical oscillator), $z_a$ ($z_m$) its position, and $Z_{a,0}$ ($Z_{m,0}$) the equilibrium position in the absence of the coupling. 
We consider the case where the coupling can be described by a potential $U_c[z_m-z_a]$ that depends only on the relative distance $z_m-z_a$ between the two oscillators (i.e.\ $U_c$ is translationally invariant). 
For example, the Coulomb interaction between two charged oscillators or the magnetic interaction between two oscillators with a magnetic moment is of this kind. 
Because $U_c$ will in general shift the equilibrium positions (see below), we define the excursions $\delta z_i = z_i - Z_i$ ($i=a,m$) with respect to the equilibrium positions $Z_i$ in the presence of $U_c$. For small excursions $\delta z_i \ll d$, where $d=Z_m-Z_a$ is the equilibrium distance, $U_c$ can be expanded as
\begin{equation}\label{eq:UC}
U_c[d+\delta z_m-\delta z_a]\approx U_c[d] + U_c'[d](\delta z_m-\delta z_a) + \frac{1}{2}U_c''[d](\delta z_m^2-2\delta z_m\delta z_a+\delta z_a^2).
\end{equation}
The coupling potential adds to the potentials of the atom and the mechanical oscillator and gives rise both to a deformation of the trap and to the desired mechanical coupling.

\subsection{Trap deformation}\label{sec:TrapDeform}
The first term in Eq.~(\ref{eq:UC}) introduces an energy offset and does not influence the dynamics.
The second term, proportional to the gradient $U_c'$, is responsible for the shift of the equilibrium positions from $Z_{i,0}$ to
\begin{equation}\label{eq:trapshift}
Z_m  = Z_{m,0}-\frac{U_c'[d]}{M\omega_{m,0}^2} \quad \textrm{and} \quad Z_a  = Z_{a,0}+\frac{U_c'[d]}{m\omega_{a,0}^2}.
\end{equation}
Because of the inverse depencence on the mass, the position shift of the mechanical oscillator is a factor $m/M \ll 1$ smaller than for the atom and can be neglected in most cases. For the atom, the shift should be small compared to the trap size in order not to expel the atom from the trap. 

The third term of Eq.~\ref{eq:UC} is the most important one and has three contributions:
The two terms quadratic in $\delta z_i$ change the trap frequencies from the unperturbed values $\omega_{i,0}$ to
\begin{equation}\label{eq:freqshift}
 \omega_m^2 = \omega_{m,0}^2+\frac{U_c''[d]}{M}  \quad \textrm{and} \quad \omega_a^2 = \omega_{a,0}^2+\frac{U_c''[d]}{m}.
\end{equation}
The frequency shift of the mechanical oscillator is suppressed again by the factor $m/M \ll 1$ compared to the atom, and is usually negligible. The change of the atomic trap frequency, on the other hand, can be substantial. In the extreme case where $U_c'' \gg m \omega_{a,0}^2$, the trap is entirely generated by $U_c$ and the trap frequency is given by $\omega_a^2 \simeq U_c''/m$. If, on the other hand, $U_c'' <0$, the trap vanishes for $|U_c''| \geq m \omega_{a,0}^2$, see Fig.~\ref{fig:CouplingPot} (b). 

If higher derivatives of $U_c$ do not vanish, the coupling potential introduces anharmonicity to the atomic trapping potential. This results in a dependence of the oscillation frequency $\omega_\mathrm{osc}$ on the atomic oscillation amplitude $a$, which can be well approximated by \cite{Antezza04}
\begin{equation}\label{eq:AnharmAnt}
\omega_\mathrm{osc}^2(a)\approx\omega_a^2 + \frac{a^2 U_c''''[d]}{8m}
\end{equation}
for small amplitudes.
Anharmonicity leads to dephasing and excitation of higher modes, especially for collective oscillations of atomic ensembles \cite{Ott03}.

\subsection{The coupling term}
The remaining mixed term in Eq.~\ref{eq:UC} gives rise to the coupling and can be interpreted as the interaction part of the Hamiltonian,
\begin{equation}\label{eq:Hint}
H_\mathrm{int}=U_c''[d]\delta z_a \delta z_m = \epsilon m \omega_a^2 \delta z_a \delta z_m.
\end{equation}

It is important to realize that the coupling scales with the same curvature $U_c''$ that also causes the change in trap frequencies in Eq.~(\ref{eq:freqshift}). 
We therefore introduce the coupling strength parameter
\begin{equation}\label{eq:Epsilon}
\epsilon = \frac{U_c''[d]}{m\omega_a^2} = \frac{U_c''[d]}{m\omega_{a,0}^2+U_c''[d]},
\end{equation}
which is the ratio of $U_c''$ and the curvature of the overall atomic trapping potential. 

We now quantize the system by promoting the amplitudes $\delta z_i$ to operators $\delta\hat{ z}_a=a_\mathrm{qm}(\hat{a}^{\dagger}+\hat{a})$ and $\delta\hat{z}_m= b_\mathrm{qm}(\hat{b}^{\dagger}+\hat{b})$, with the ground state amplitudes $a_\mathrm{qm}=\sqrt{\hbar/2m\omega_a}$ and $b_\mathrm{qm}=\sqrt{\hbar/2M\omega_m}$ and the bosonic creation and annihilation operators satisfying the usual commutation relations $[\hat{a},\hat{a}^{\dagger}]=1$ and $[\hat{b},\hat{b}^{\dagger}]=1$. 
This casts Eq.~(\ref{eq:Hint}) into the form
\begin{equation}\label{eq:Hqm}
\hat{H}_\mathrm{int} =  \epsilon m\omega_a^2 \delta \hat{z}_a \delta\hat{z}_m = 
\epsilon\frac{\hbar\omega_a}{2}\sqrt{\frac{\omega_a}{\omega_m}}\sqrt{\frac{m}{M}} (\hat{a}^{\dagger}+\hat{a})(\hat{b}^{\dagger}+\hat{b}).
\end{equation}
This is a linear coupling in the amplitude, which simplifies for near-resonant coupling $\omega_a \simeq \omega_m$ and in the rotating wave approximation to
\begin{equation}
\hat{H}_\mathrm{int}=\hbar g_0 (\hat{a}^{\dagger}\hat{b}+\hat{a}\hat{b}^{\dagger})
\end{equation}
with the single-phonon coupling constant
\begin{equation}\label{eq:Coupling}
g_0 = \epsilon \frac{ \omega_a}{2}\sqrt{\frac{m}{M}}.
\end{equation}

The linear coupling is in contrast to the dispersive opto- or electromechanical schemes, where the coupling is quadratic in the cavity field amplitude $\hat{c}$, $\hat{H}^\mathrm{om}_\mathrm{int}=\hbar g^\mathrm{om} \hat{c}^{\dagger}\hat{c}(\hat{b}^{\dagger}+\hat{b})$. In optomechanics, a large frequency difference between the cavity and the mechanical oscillator can be tolerated, since the cavity can be strongly driven and the linearized dynamics in a frame rotating at the drive frequency is considered. For the direct linear coupling between two mechanical oscillators considered here (undriven or only weakly driven), significant energy transfer occurs only near resonance.

Let us discuss some of the implications of the obtained coupling constant.
It contains a disadvantageous term $\sqrt{m/M}\ll 1$ that reflects the impedance mismatch between the two oscillators. For a single atom with $m\sim10^{-22}$~g and micro- or nanostructured oscillators with $M\sim10^{-13}-10^{-7}$~g \cite{Favero07,Verbridge08,Thompson08}, we obtain $\sqrt{m/M} = 10^{-8} - 10^{-5}$, resulting in a strong reduction of the coupling rate.
Next, $g_0$ scales with the resonance frequency, which is limited by the experimentally achievable curvature of the atomic trapping potential. While magnetic microtraps and optical lattices can reach $\omega_a/2\pi \sim1~$MHz \cite{Reichel02,Gehr10}, ion traps have reached up to 50~MHz \cite{Jefferts95}. 
Finally, $g_0$ contains the coupling strength parameter $\epsilon$ as defined in Eq.~(\ref{eq:Epsilon}). 
In most cases, this parameter will be small, $\epsilon \ll 1$, to avoid strong distortion of the trapping potential by $U_c''$. If on the other hand the trap is designed such that $U_c''$ itself provides the atomic confinement, the parameter saturates at $\epsilon = 1$. 
It is thus not possible to increase $\epsilon$ at will by simply increasing $U_c''$. 
For $U_c''<0$ it is actually possible to reach $|\epsilon|>1$, but in this regime $\omega_a$ decreases with increasing $|U_c''|$ and the trap becomes very shallow and strongly distorted, see Fig.~\ref{fig:CouplingPot} (b). As long as no other measures such as compensation of the trap distortion are taken (see below), we thus have
\begin{equation}\label{eq:CouplingCondition}
\epsilon \lesssim 1.
\end{equation}
Overall, $g_0 \lesssim 10^{-5}\, \omega_a$ is typically in the regime of $g_0/2\pi=10^1-10^3$~Hz for the direct mechanical coupling of a single atom or ion to a nanostructured oscillator.

\subsection{Dissipation}
To put the coupling rate into context, one has to compare it to the dissipation rates in the system.
There, the question about the mechanical quality factor of the atom arises. Since levitated systems such as ultracold atoms have no direct contact with the environment, there are only minute damping mechanisms present (e.g. the friction caused by collisions with the background gas, by resistive damping of moving mirror charges in an ion trap, or by damping of induced currents from moving paramagnetic atoms). 
Larger effects on the motion come from trap anharmonicity and residual heating due to noise in the structures generating the trap (fluctuating currents, voltages, magnetization). In particular, slow drifts of trap parameters lead to drifts of the resonance frequency and cause dephasing. While experiments with collective motion of ultracold atoms have achieved quality factors $Q\sim10^4$ \cite{Harber05} limited by the atomic trap lifetime and rather low trap frequencies, trapped ions can achieve $Q\sim10^{10}$ in cryogenic Penning traps \cite{Haeffner00,Verdu04}. However, $Q$ is in many cases not the relevant parameter, and particularly in the context of quantum engineering, the decoherence rate $\gamma_\mathrm{dec}$ is the key figure of merit. In ion traps, nonclassical motional states with coherence lifetimes up to $180$~ms have been reported \cite{Lucas07,Jost09}.
For neutral atoms, nontrivial motional quantum states have been studied in optical lattices by populating excited lattice bands \cite{Bouchoule99,HeckerDenschlag02,Spielman06,Mueller07,Foerster09},
and in magnetic microtraps in the context of trapped atom interferometers \cite{Boehi09,Riedel10,Jo07,Schumm05,Cronin09}, where decoherence lifetimes up to 200~ms have been demonstrated \cite{Jo07}.
Furthermore, for some systems it is possible to map motional states to internal states which have much longer coherence lifetimes. This could be used to preserve quantum states for several seconds.

On the other hand, decoherence of the mechanical oscillator is typically fast despite a high mechanical quality factor, since the coupling to the environment introduces a dependence on the temperature $T$ of the bath,
\begin{equation}
\gamma_\mathrm{m,dec}=\frac{k_B T}{\hbar Q}.
\end{equation}
For standard cryogenic cooling with $T=4~$K and good quality factors $Q=10^5$ one ends up with $\gamma_\mathrm{m,dec}=2\pi\times1~$MHz. However, for state of the art cryogenics with $T=10~$mK and exceptional quality factors $Q=10^7$ as demonstrated in \cite{Zwickl08} one may achieve $\gamma_\mathrm{m,dec}=2\pi\times20~$Hz, sufficient to achieve the strong coupling regime where $g_0 > (\gamma_\mathrm{m,dec}, \gamma_\mathrm{a,dec}$). While the large mechanical decoherence rate is a general challenge for all experiments aiming at the observation of quantum behaviour of mechanical oscillators, it is particularly severe in the presence of low coupling rates.
Strategies to increase the coupling beyond the values given above are thus highly desired.

\subsection{Strategies for stronger coupling} 
A natural way to increase the coupling rate is to couple many atoms simultaneously to the oscillator. Coupling to the atomic center of mass mode $\hat{a}=\sum_i \hat{a}_i/\sqrt{N}$ of an ensemble of $N$ atoms, one obtains a collectively enhanced coupling strength $g_N = \sqrt{N}g_0$. We note that this result can be obtained by simply replacing $m$ with the mass $Nm$ of the whole ensemble in Eq.~(\ref{eq:Coupling}). The coupling rate increases with $\sqrt{N}$, and for large atomic ensembles with $N\sim 10^8$, a collectively enhanced coupling of $g_N \simeq 10^{-2}\, \omega_a$ could be reached. A system benefitting from this approach is discussed in Sec.~\ref{sec:LatticeCoupling}.

Furthermore, the coupling strength parameter $\epsilon$ can be increased beyond the limit given in Eq.~(\ref{eq:CouplingCondition}) if the trap deformation is compensated with the help of additional potentials. This could allow one to increase $g_0$ by about two orders of magnitude. We give an example in Sec.~\ref{sec:IonCoupling}.

An alternative approach is to focus on molecular-scale oscillators such as carbon nanotubes. Due to the ultimately low effective mass $M\sim10^{-16}$~g, both $g_0$ and $g_N$ can be increased by about two orders of magnitude compared to the more massive oscillators discussed above. This scenario is further discussed in Sec.~\ref{sec:CNTCoupling}.

Finally, the coupling can be significantly enhanced with the help of high-finesse optical cavities. The cavity acts as a lever and provides a coupling that is independent of the mass ratio. This is possible because the cavity breaks the translational invariance, i.e.\ the interaction potential is no longer of the form $U_c[z_m-z_a]$ considered above. This scheme provides much stronger coupling of up to  $g_0 = 3\times10^{-2}\, \omega_a$ and we discuss it in Sec. \ref{sec:CavityCoupling}.

We point out that even if the absolute coupling rate is moderate, the normalized coupling rate, i.e.~$g_0$ compared to the oscillator frequency, is rather large for coupled atom-oscillator systems. In optical cavity QED experiments, by comparison, the highest normalized coupling rates achieved so far amount to $g_0/\omega_{cav}=6\times10^{-7}$ \cite{Colombe07}, with $\omega_{cav}$ being the optical resonance frequency of the cavity. Only recently, solid-state based cavity QED experiments in the microwave regime could enter the ultrastrong coupling limit with larger normalized coupling strengths of up to $g_0/\omega_{cav}=1\times10^{-1}$ \cite{Niemczyk10}. In this regime, the counter-rotating terms of the coupling Hamiltonian become relevant and lead to new effects, which would be interesting to study for mechanical oscillators. Remarkably, in the first experiment demonstrating strong coupling of nanomechanical motion to a superconducting qubit, $g_0/\omega_m=1\times10^{-2}$ was achieved \cite{OConnell10}.

\section{Coupling a single ion to a mechanical oscillator}\label{sec:IonCoupling}
Trapped ions are one of the best controlled systems with elaborate techniques for quantum state engineering. In particular, ion motion can be controlled on the single quantum level by addressing motional sidebands of optical transitions with lasers \cite{Meekhof96,Jost09}, and quantum states of motion are routinely employed in experiments e.g.\ on quantum computing \cite{Benhelm08,Monz09}. Coherence of motional quantum states has been maintained for more than hundred milliseconds \cite{Lucas07}, and thus, ions are a particularly promising coupling partner for micro- and nanomechanical oscillators.
A further advantage is the availability of a strong coupling force: An electrically charged mechanical oscillator can exert a substantial force on the ion, allowing to couple the mechanical motion to the ion motion. Several proposals \cite{Wineland98,Tian04,Hensinger05} have considered such a coupling.
Notably, this coupling has been recently demonstrated at the single quantum level for an ion and the smallest mechanical oscillator available, namely another ion \cite{Brown10}.

We want to estimate the achievable coupling rate for an ion-oscillator system similar to Ref.~\cite{Wineland98}. The coupling constant can be written in the form of Eq.~(\ref{eq:Coupling}), where the details about the coupling potential are absorbed in $\epsilon$.
More explicitly, a charged nanoscale oscillator would provide a coupling via the Coulomb potential $U_c[d]=1/(4\pi\epsilon_0)\times eq/d$, where $e$ is the elementary charge of a singly charged ion, and $q$ the charge on the tip of the oscillator.
As discussed above, in order to avoid a strong distortion of the trap, the position shift of the ion due to the gradient $U_c'[d]$ has to be small compared to $d$, which yields the condition $eq/4\pi\epsilon_0 d^3 m\omega_{a,0}^2 \lesssim 1$, and the curvature $U_c''[d]$ has to be smaller than the curvature of the ion trap, or equivalently $\epsilon=eq/2\pi\epsilon_0 d^3 m\omega_{a}^2 \lesssim 1$, giving essentially the same limit.
While the coupling rate scales with $\omega_a \sqrt{m}$, the maximum achievable trap frequency in an ion trap scales with $1/m$, such that lighter ions achieve a larger coupling. For a $^9$Be$^+$ ion in a trap with $\omega_a=2\pi\times70~$MHz and a low-mass oscillator with $M=10^{-12}$~g one obtains a coupling constant $g_0\simeq 2\pi\times 150~$Hz for $\epsilon\simeq1$.
One can see that even though the high achievable trap frequency is advantageous, the coupling remains modest. To achieve $\epsilon\simeq 1$ for this case, we assume a nanoscale oscillator with a conductive sphere at the tip with a capacity $C=4\pi\epsilon_0 r$, proportional to the sphere radius $r$. For $\epsilon=1$ one has to charge the sphere with a voltage $V = m\omega^2d^3/2re$, which for $d=10~\mu$m and $r=100~$nm amounts to $V=90~$V.

To benefit from the excellent motional control, coupling to a single or very few ions is preferable, and collective enhancement is probably not the way to go. To achieve stronger coupling one can increase $U_c''[d]$ further, e.g.\ by decreasing $d$ or increasing $V$, if one compensates for the resulting distortion of the ion trap.
Additional trap electrodes, not mounted on the oscillator, allow one to compensate the static electric field, gradient, curvature, and perhaps also higher-order derivatives of $U_c$, while the oscillating component is not reduced. With this approach, the distorting effect of the coupling force can be reduced and $\epsilon$ can be made larger than unity.
Note that this requires precisely tuned compensation voltages about a factor $\epsilon$ larger than the voltages that produce the unperturbed trap, and also the voltage charging the oscillator has to be larger by the same factor. This is experimentally challenging, and compensation may allow one to increase the coupling by 1-2 orders of magnitude, such that $\epsilon\lesssim 10^2$. 
Similar compensation techniques could also be employed in the case of magnetic coupling, see Sec.~\ref{sec:MagneticCoupling}.

\section{Coupling a BEC to a mechanical oscillator via surface forces}\label{sec:SurfaceCoupling}

\begin{figure}[t]
\centering
\includegraphics[width=0.36\textwidth]{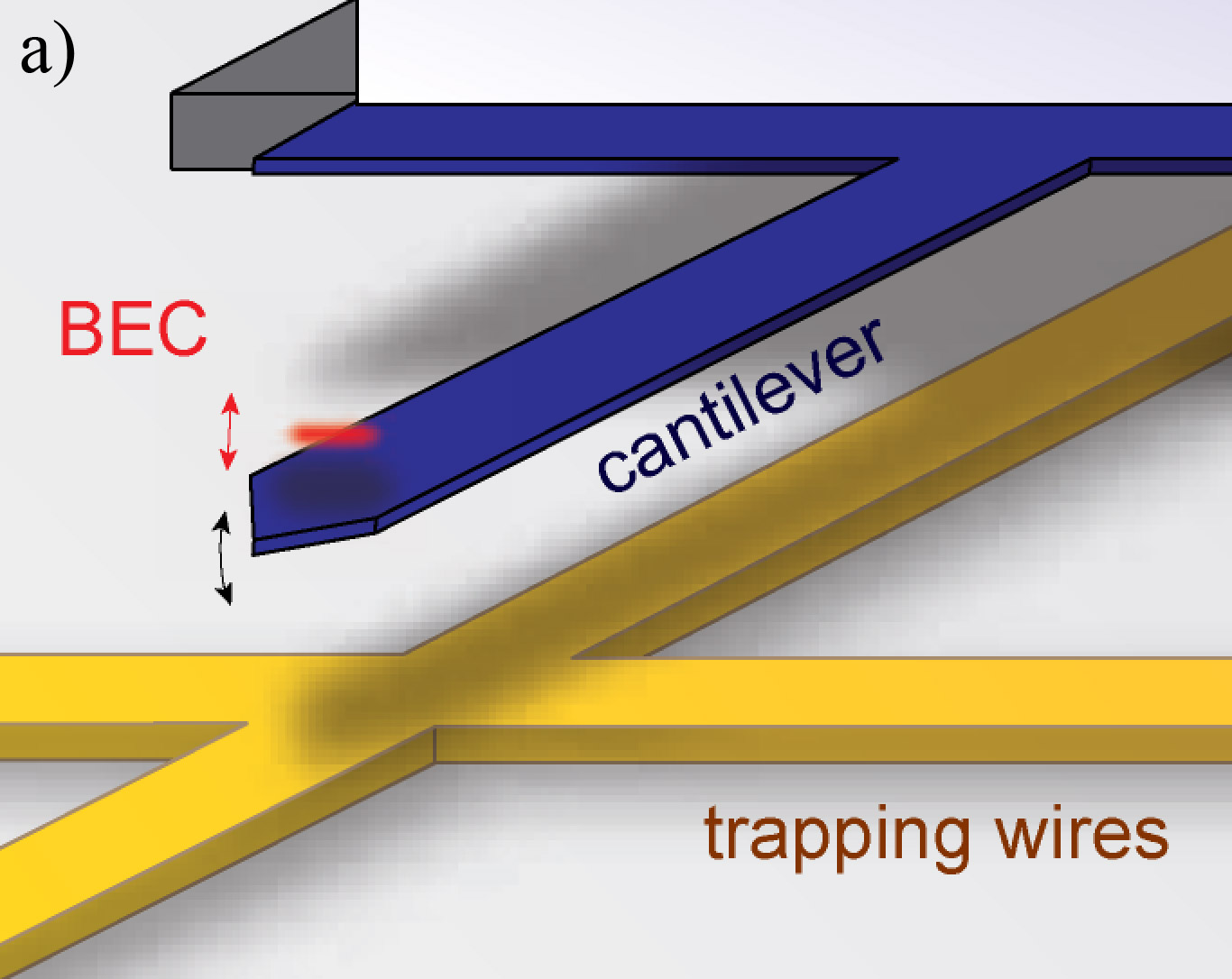}\includegraphics[width=0.60\textwidth]{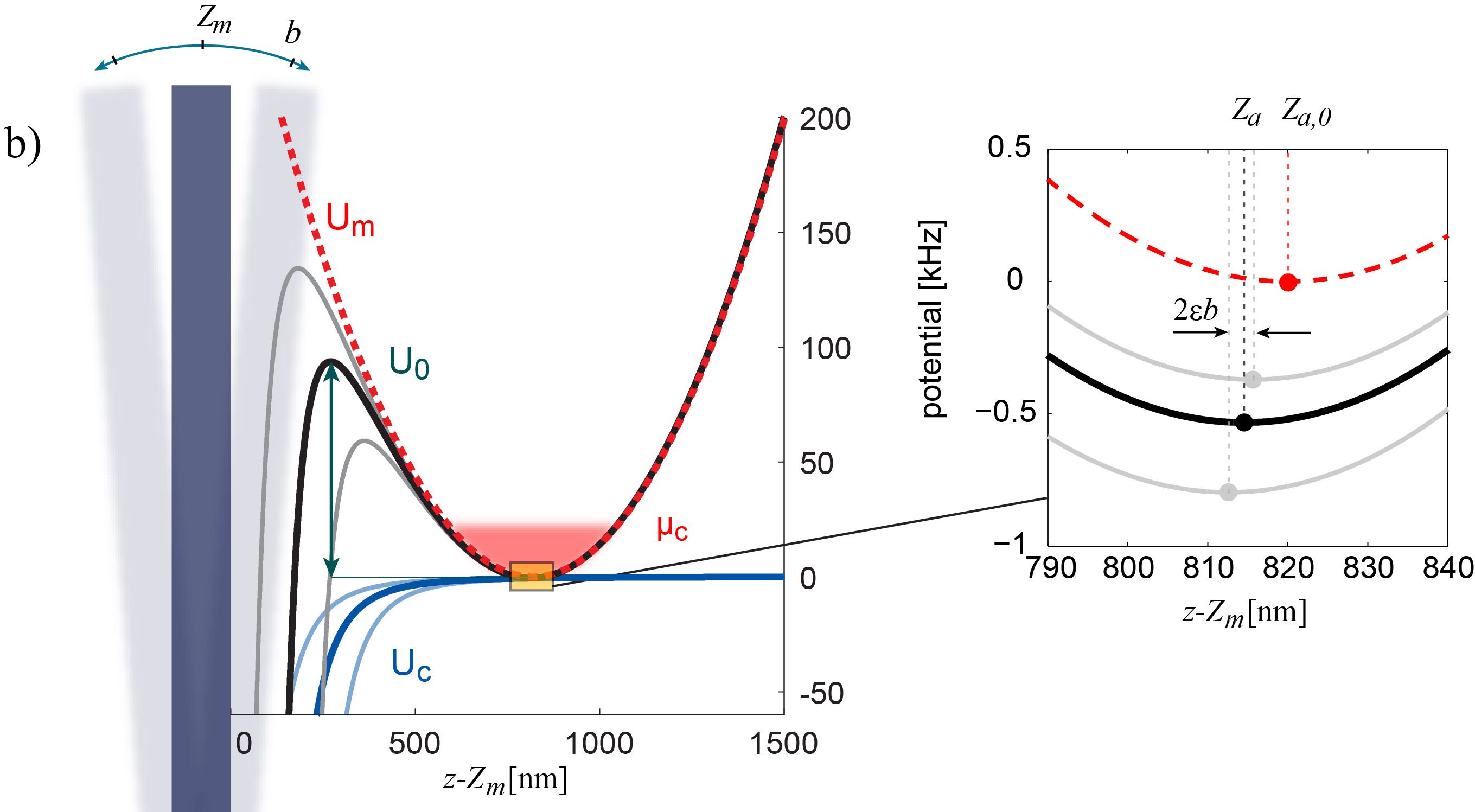}
\caption{\label{fig:SurfForceSchem} a) Schematic overview of the experiment. A micromechanical cantilever is integrated on an atom chip with microfabricated gold wires for magnetic trapping. A BEC is created and positioned close to the cantilever in order to sense mechanical oscillations. Figure adapted from \cite{Hunger10a}. b) Calculation of the undisturbed magnetic potential $U_m$ (red, dashed) and the resulting trapping potential (black solid line) destorted by the {C}asimir-{P}older potential $U_c=\UCP$ (blue, solid) for small atom-surface distance. It leads to a barrier of height $U_0$, a shift of the trap minimum and a change in trap frequency. Parameters are $\omega_a/2\pi=10~$kHz and $d=820~$nm. The chemical potential $\mu_c$ of a BEC with $N=600$ atoms is shown for comparison. Oscillations of the cantilever modulate the trapping potential (light blue, grey). Right panel: Enlarged view of the potential close to the trap minimum (indicated by the box). Modulation of the trap minimum position with an amplitude $\epsilon b$ leads to excitation of collective motion.
}
\end{figure}

The motion of neutral atoms can be coupled to mechanical oscillators via a distance-dependent potential, similar to the case of ions. In the following we summarize an experiment where such direct motional coupling was demonstrated for the first time \cite{Hunger10a}. The experiment studies the coupling of a mechanical oscillator to collective oscillations of a Bose-Einstein condensate (BEC) in a trap. The fact that the atoms are Bose-condensed is important because of the small spatial extension of the condensate compared to a thermal gas. While in the present experiment the quantum nature of the BEC is not directly exploited, it is an important ingredient of future work since it provides the initialization of the atoms in the motional ground state. The employed oscillator is a commercial micromechanical cantilever with dimensions $(l,w,h)=(195,35,0.4)~\mu$m, an effective mass of $M=5~$ng, a fundamental resonance frequency $\omega_m=2\pi\times 10~$kHz, and a quality factor $Q=3200$.
The coupling is realized via attractive surface forces which are of fundamental origin and do not require functionalization of the mechanical oscillator.
Surface forces typically decay over a few micrometers distance and thus require exceptional control over the position of the atoms. In the experiment, magnetic microtraps on an atom chip are used to create, hold and position BECs with a few nanometers precision close to an on-chip integrated microcantilever, see Fig. \ref{fig:SurfForceSchem} (a).

For small atom-surface distance, the magnetic potential $U_m$ used to trap the atoms is strongly deformed by the surface potential $U_c=\UCP+\Uad$, where $\UCP$ denotes the Casimir-Polder (CP) potential \cite{Casimir48}, and $\Uad$ includes possible additional potentials due to stray fields. In the direction perpendicular to the surface, the combined potential is (see Fig.~\ref{fig:SurfForceSchem} b)
\begin{eqnarray}
U[z] &=&  U_m + \UCP + \Uad + U_\mathrm{grav}  \nonumber\\
 &=& \frac{1}{2}m\omega_{a,0}^2(z-Z_{a,0})^2  -\frac{C_4}{(z-Z_m)^4} + \Uad[z-Z_m] + m g z.
\end{eqnarray}
Here, $Z_m$ is the position of the cantilever surface, $C_4$ the CP-coefficient for a perfect conductor, $Z_{a,0}$ the unperturbed minimum position of the magnetic trap, $\omega_{a,0}$ the unperturbed trap frequency, $m$ the atomic mass, and $g$ the acceleration in the gravitational potential $U_\mathrm{grav}$.

\begin{figure}
\centering
\includegraphics[width=0.77\textwidth]{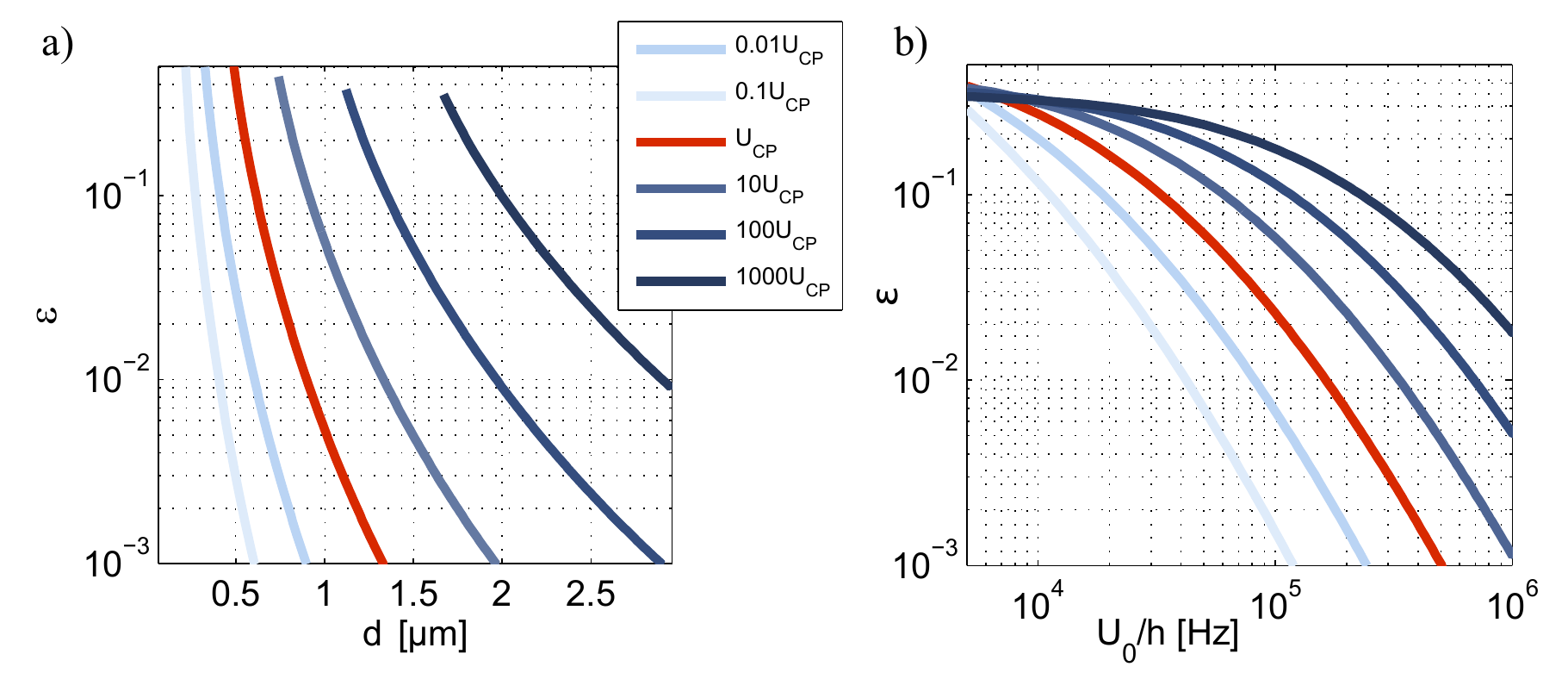}
\caption{\label{fig:EpsVsDist} (a) Coupling strength parameter $\epsilon$ for different surface potential strength $U_c=\beta\UCP$ as a function of atom-surface distance and (b) barrier height. Chosen parameters are $\omega_{a,0}=2\pi\times10~$kHz and $\UCP=-C_4/d^4$ with $C_4$ the CP-coefficient for a perfect conductor. Stronger surface potential with large $\beta$ can arise e.g. from adsorbates or surface contaminants, weaker potentials arise e.g.\ for molecular-scale oscillators.
}
\end{figure}

To estimate the achievable coupling strength for this kind of coupling potential, we calculate the coupling strength parameter  $\epsilon=20C_4/m\omega_a^2d^6$ for a pure CP potential. In many cases, additional (surface) potentials originating e.g.\ from adsorbates, stray charges, or static external charging, have a very similar distance dependence as $\UCP$. Fig.~\ref{fig:EpsVsDist} shows $\epsilon$ for the case of a harmonic trap with $\omega_{a,0}=2\pi\times10~$kHz and attractive coupling potentials of the form $U_c[d]=-\beta C_4/d^4$ for several coefficients $\beta$. One finds that the trap vanishes for small atom surface distance, where the coupling strength parameter reaches its maximum $\epsilon \simeq 0.3$. The rapid decay of $\epsilon$ with the distance can be mitigated by stronger coupling potentials, where $\epsilon>0.1$ can be achieved also for deep traps (see Fig. \ref{fig:EpsVsDist} b) with up to 10 bound levels.
For a typical experimental value $\epsilon=0.15$, the coupling rate for a single atom amounts to $g_0/2\pi = 2.5 \times 10^{-4}~$Hz in this system, too small for the coupling to be detectable on the single-phonon level. However, for a coherently driven cantilever, the atoms can be used as a probe to readout the oscillator vibrations. 

\subsection{Experimental Results}
\begin{figure}[t]
\centering
\includegraphics[width=0.59\textwidth]{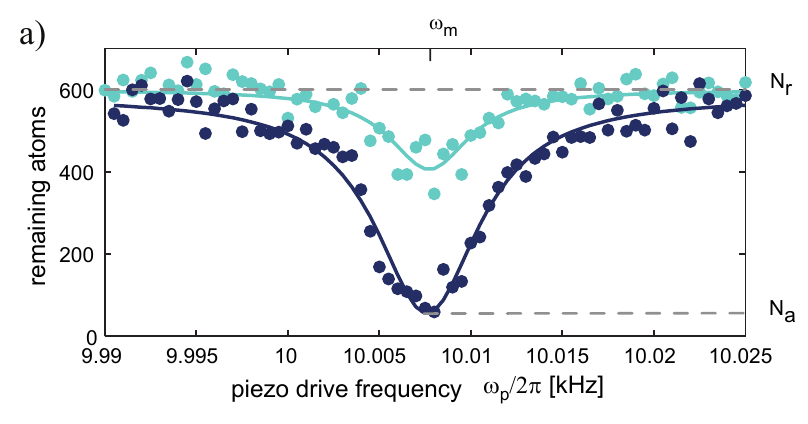}
\includegraphics[width=0.39\textwidth]{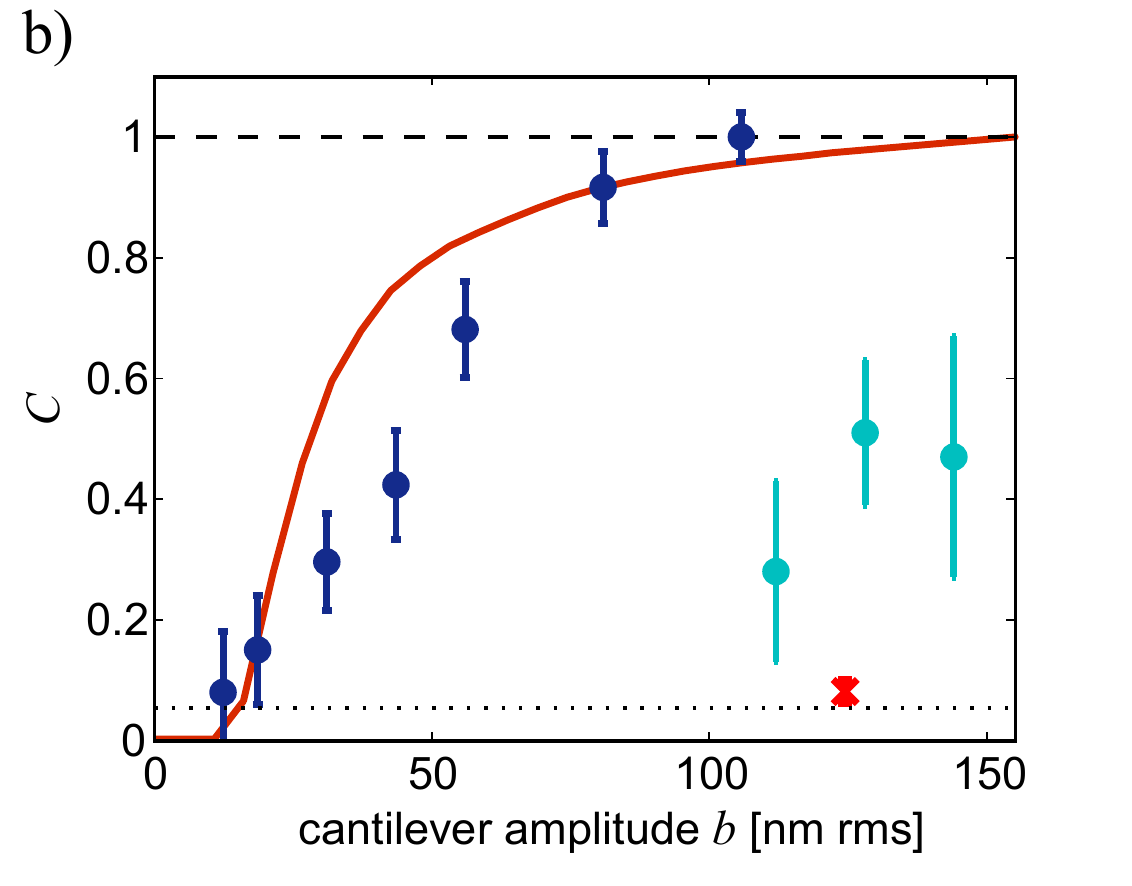}
\caption{\label{fig:DynLoss} Remaining atom number in a resonant trap for varying cantilever excitation frequency $\omega_p$. The dark (light) blue circles correspond to a cantilever amplitude $b=120~$nm ($50~$nm) on resonance. Solid lines: Lorentzian fits with $6~$Hz FWHM, corresponding to the width of the cantilever resonance. The remaining atom number with ($N_a$) and without ($N_r$) resonant piezo excitation of the cantilever is indicated. Figure adapted from \cite{Hunger10a}.
b) Contrast $C$ for the c.o.m.\ mode as a function of the cantilever amplitude $b$. For measurements on the metallized side (dark blue) a $3.2\pm0.6$ times larger coupling signal than on the dielectric backside (light blue) is found. For comparison, the contrast for an off-resonant trap with $\omega_a/2\pi=4~$kHz is shown (red). The dotted line indicates the rms noise of the measurement. The red solid line is a 1D numerical simulation with no free parameters.
}
\end{figure}

In the experiment, coherently driven cantilever oscillations are coupled to the motion of trapped ultracold atoms nearby. The atoms are ramped to a typical distance of $d=1.5~\mu$m from the cantilever, where the magnetic trap is markedly affected by the static surface potential and one reaches values of $\epsilon=0.15$ for a trap depth $U_0= 8\times \hbar \omega_a~$. To achieve resonant interaction, the trap frequency is tuned to the eigenfrequency of the cantilever's fundamental mode.
When the cantilever is excited with a piezotransducer at different frequencies around its resonance, mechanical oscillations excite the center of mass (c.o.m.) mode of the condensate, leading to trap loss over the potential barrier. In this way, the resonance spectrum of the cantilevers fundamental mode can be probed by the atoms (see Fig.~\ref{fig:DynLoss} (a)).

The sensitivity of this readout method can be inferred by measuring the coupling signal on resonance as a function of the cantilever amplitude. Such measurements can be performed both on the metallized frontside and on the dielectric backside of the cantilever.
Figure \ref{fig:DynLoss} b) shows measurements of the contrast $C=(N_r-N_a)/N_r$ for the c.o.m.\ mode with $N_r$ ($N_a$) the remaining atom number off (on) resonance as a function of the cantilever amplitude. For comparison, also the signal for an off-resonant trap with $\omega_a/2\pi=4~$kHz is shown. The measurements yield a minimum resolvable r.m.s.\ cantilever amplitude of $b_\mathrm{rms}=13\pm4~$nm for unity signal to noise ratio without averaging. The contrast for a given amplitude is different for the front and the backside of the cantilever. With additional measurements and simulations, it is possible to quantify the difference in the strength of the surface potential (see also Fig. \ref{fig:CNTCoupStr}). It was found that on the metallized side $U_s=(200\pm100)\times C_4/d^4$ while on the dielectric side it is consistent with a CP potential within the errorbars. A likely origin of the stronger potential is the presence of surface adsorbates which cause electric stray fields \cite{Obrecht07,Hunger10b}. The potential strength can be quantitatively reproduced when considering the BECs deposited on the surface during operation of the experiment. The larger electric dipole moment of adsorbates and the larger number of experiments performed on the metallized side are consistent with the stronger potential.

Two effects influence the observed sensitivity. First, the finite lifetime of the atoms in the trap due to parasitic loss limits the coupling duration and thus the achievable excitation of the atoms for small amplitudes. Second, trap anharmonicity leads to dephasing of the cloud oscillations and thereby to a maximum cloud amplitude for a given cantilever amplitude. More insight into the dynamics of the cloud and the influence of the trap anharmonicity can be obtained from a 1D numerical integration of the time dependent Gross-Pitaevskii equation. The measured data is well reproduced by the simulation where only the inferred strength of the surface potential and measured parameters of the experiment enter (Fig.~\ref{fig:DynLoss} b).

In the measurements, trap loss was used as the simplest way to detect the BEC dynamics induced by the coupling. This implies that in order to be observable, the excitation has to lead to large amplitude collective oscillations, such that a part of the cloud spills over the barrier and is lost from the trap. By contrast, BEC amplitudes down to the single phonon level could be observed by direct imaging of the motion.
A coherent state $| \alpha \rangle$ of the c.o.m.\ mode of $N=100$ atoms with $\alpha =1$ released from a relaxed detection trap with $\omega_a=2\pi\times 100$~Hz has an amplitude of $\sqrt{2\hbar\omega_a/mN} \alpha t=400$~nm after $4$~ms time-of-flight. This is about $10\%$ of the BEC radius and could be resolved by absorption imaging with improved spatial resolution.
From the numerical simulation of the cloud excitation we estimate that $b_\mathrm{rms}=0.2$~nm would excite the BEC to $\alpha =1$ within an interaction time of $20$~ms and could thus be detected. This should allow detection of the thermal amplitude $a_{th}=\sqrt{k_BT/M\omega_m^2}=0.4~$nm of the used cantilever.

\begin{figure}[t]
\centering
\includegraphics[width=0.65\textwidth]{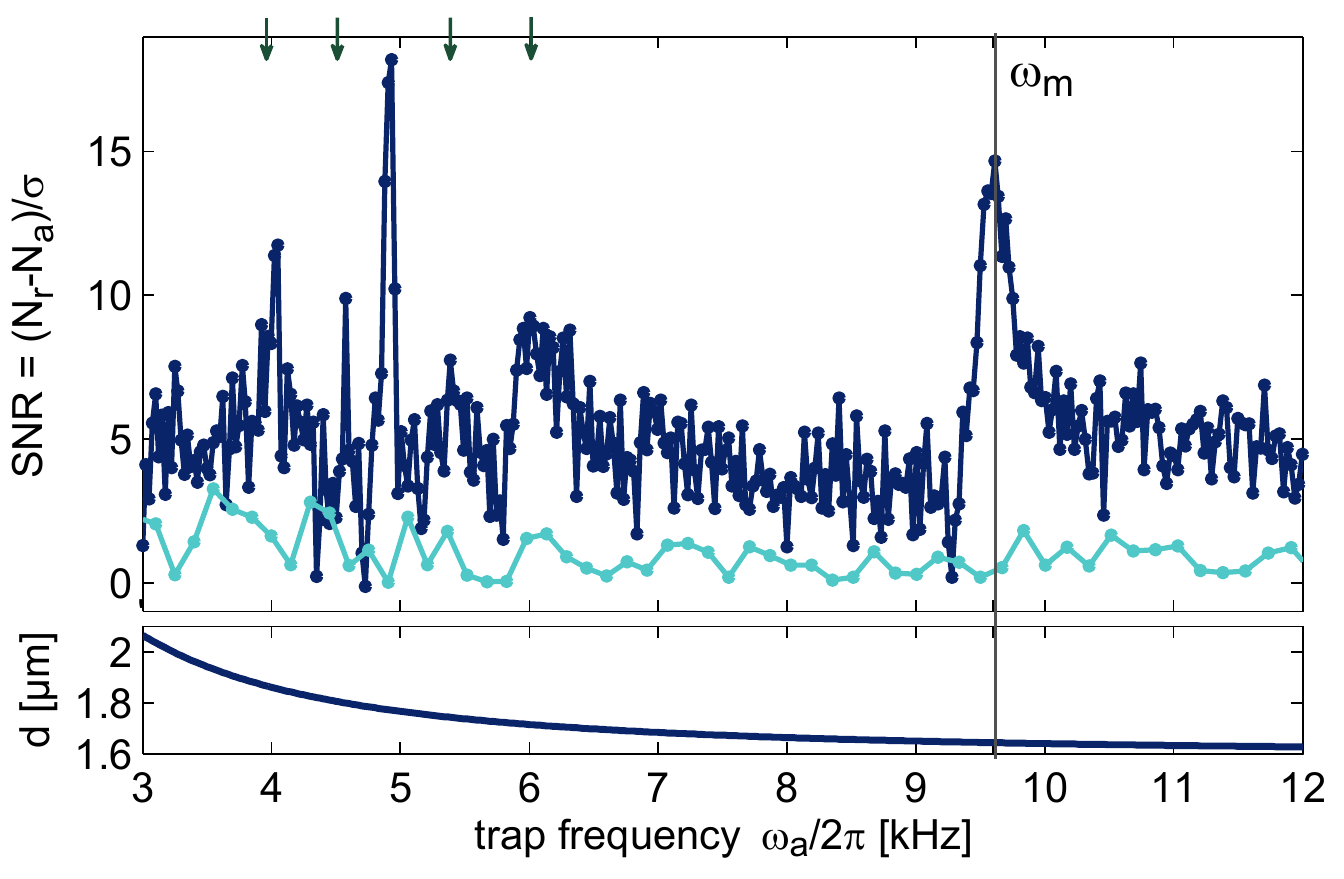}
\caption{\label{fig:Cantres} Top graph: BEC response as a function of $\omega_a$, for fixed resonant piezo excitation ($b=180~$nm) (dark blue). Datapoints are connected with a line to guide the eye. Light blue: reference measurement without piezo excitation. Two major resonances at $\omega_m=\omega_a$ (c.o.m. mode) and $\omega_m=2 \omega_a$ (parametric breathing mode) and up to four smaller ones (green arrows) are observed. Due to cantilever aging, $\omega_m/2\pi=9.68~$kHz in this measurement. Bottom graph: Set values of $d$, chosen to keep $N_r$ sufficiently constant and to avoid saturation of $N_a$. Figure adapted from \cite{Hunger10a}.}
\end{figure}

So far, the described measurements were performed for resonant coupling, where the trap frequency and the mechanical resonance frequency match. Both single atoms as well as ensembles of atoms furthermore show parametric resonances, and the latter also exhibit shape resonances. The spectrum of collective mechanical modes is influenced by atomic collisions, which leads to additional, shifted resonance frequencies compared to the harmonic spectrum of a non-interacting gas \cite{Stringari96,Kimura99}. When anharmonicity is present, resonance broadening and nonlinear mode mixing can further modify the spectrum.
Studying the atomic mode spectrum can give insight about which modes can be excited efficiently, how sensitive the atomic response depends on the trap frequency, and to which extent the strong anharmonicity of the coupling trap affects the situation.
With a measurement of the dependence of the atomic response on the trap frequency $\omega_a$, a spectroscopy of the strongest cloud excitations can be performed. It is found that the cantilever can be coupled selectively to different, spectrally narrow BEC modes, see Fig. \ref{fig:Cantres}. The sharp resonances indicate coherent excitation and provide a means to control the coupling by changing the trap frequency.

\subsection{Coupling a BEC to a carbon nanotube}\label{sec:CNTCoupling}
Coupling ultracold atoms and mechanical oscillators via surface forces has a central advantage: The employed coupling force is of fundamental origin, and functionalization of the oscillator, e.g.\ by the fabrication of mirrors, magnets or electrodes on the oscillator tip, is not required. This is 
in contrast to most of the theoretical proposals considering atom-resonator coupling so far.
Coupling via surface forces could thus be used to couple atoms to molecular-scale oscillators.
Due to their ultimately low mass and the high achievable quality factor, single wall carbon nanotubes (CNT) are particularly promising in this context.

Considering a suitable geometry with a nanotube diameter $d_\mathrm{CNT} = 1.5~$nm and length $l=15~\mu$m we obtain $\omega_m=2\pi\times 20~$kHz and an effective mass of $M=2\times10^{-20}~$kg. This leads to a room temperature thermal motion amplitude of $b_{th}=\sqrt{k_BT/M\omega_m^2}= 4~\mu\mathrm{m}$, much larger than typical atom-surface distances in the previously described experiment.
Even more impressive, the ground state amplitude is almost macroscopic, $b_\mathrm{qm}=\sqrt{\hbar/2M\omega_m}=0.2~\mathrm{nm}$. These numbers are about four orders of magnitude larger than those of the cantilever used in the experiment above. If the readout scheme of direct imaging of the cloud excitation is employed, the expected sensitivity should permit measurements close to the ground state of the nanotube.

However, one central concern is whether the strength of the surface potential of such nanoscale objects is sufficient for coupling experiments.
From an estimation of the Casimir-Polder potential of a CNT \cite{Fermani07} we infer a potential $U_\mathrm{CP,CNT}\approx 0.06\times\UCP$ in the relevant distance range, where $\UCP$ is the potential of a bulk conductor. The achievable coupling strength parameter for a resonant trap with sufficient trap depth in this case is of the order $\epsilon=1\times10^{-2}$.

\begin{figure}[t]
\centering
\includegraphics[width=0.55\textwidth]{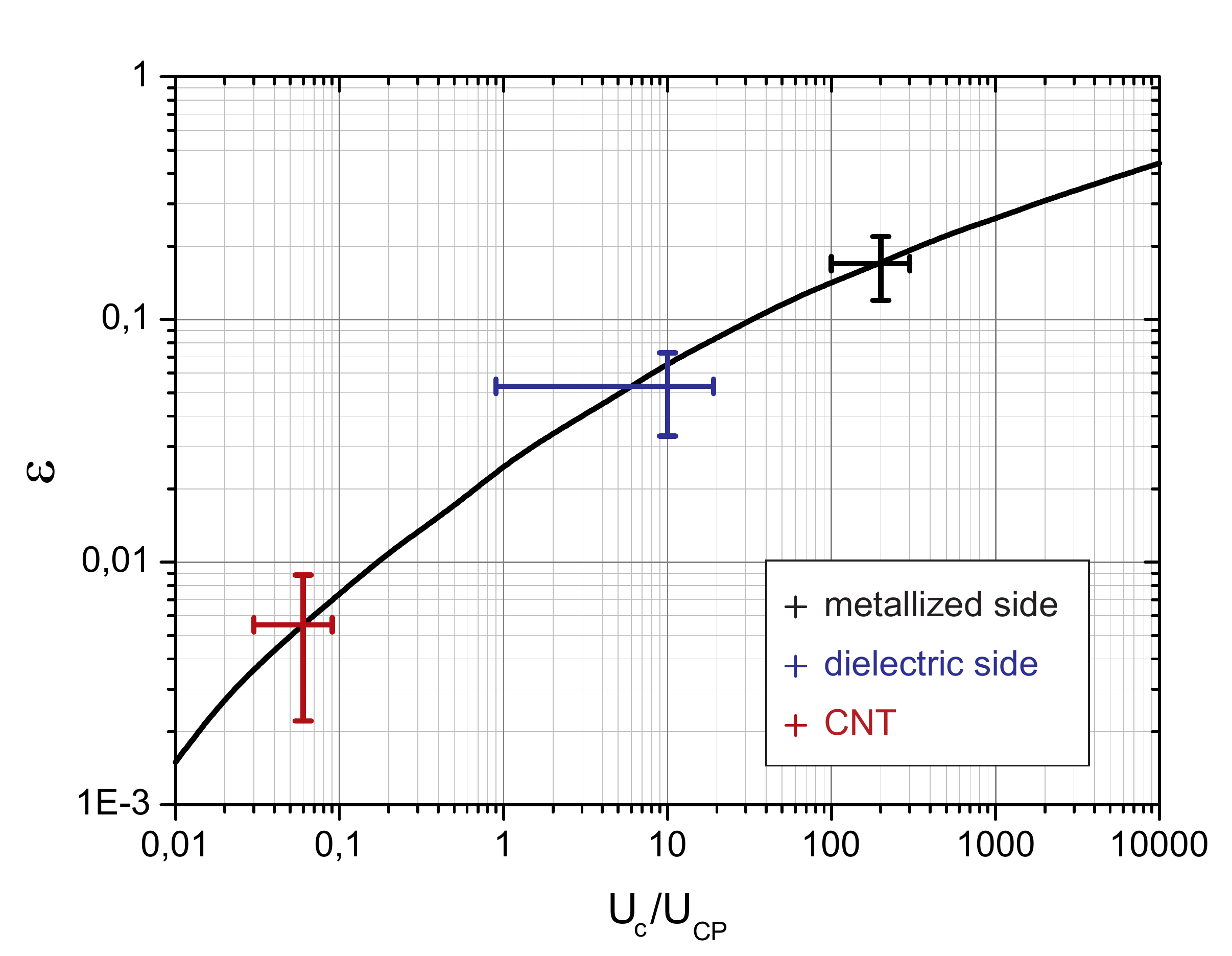}
\caption{\label{fig:CNTCoupStr} Variation of the coupling strength parameter $\epsilon$ as a function of surface potential strength normalized to the CP potential of a perfect conductor, calculated for a trap depth of $U_0=8\times \hbar \omega_a$ (solid line). In addition, the evaluated $\epsilon$ above the metallized and dielectric side of the cantilever used in the experiments and the expected value for a CNT are shown.}
\end{figure}

Stronger coupling could be realized e.g.\ by static charging of the CNT. Electric fields $E[d]$ polarize neutral atoms via their static polarizability $\alpha_0$, and a potential $U_c[d]=-(\alpha_0/2)|E[d]|^2 = -(\alpha_0/2)(q/4\pi\epsilon_0d^2)^2\propto -1/d^4$ arises, with the same distance dependence as the CP potential. Due to the small atom-surface distance, the electric field and also the field gradient will be large even for small charging voltages. Since CNTs can be fabricated in metallic or semiconducting configuration, charging does not require deposition of electrodes on the nanotube, and electric contacting is a standard technique that is compatible with high quality factors \cite{Huettel09}.
By charging the nanotube, a potential much stronger than the CP potential could be generated, and coupling strength parameters of $\epsilon\simeq 0.3$ might be possible. Figure \ref{fig:CNTCoupStr} shows the coupling strength parameter $\epsilon$ as a function of the strength of the surface potential. The observed $\epsilon$ at the both sides of the cantilever used in the described experiment is shown together with the expected value for a CNT.

Finally, coupling schemes that do not rely on surface forces could be employed. One possibility is to use a doubly clamped, suspended CNT as current carrying wire to generate a magnetic trap. Several papers have investigated this situation theoretically \cite{Peano05,Fermani07,Petrov09} and studied its feasibility for atom trapping. In this configuration, mechanical oscillations of the nanotube translate directly into trap oscillations and provide a coupling strength parameter $\epsilon=1$. In combination with compensating fields, even values of $\epsilon>1$ could be possible.

We now estimate the achievable coupling parameters for a CNT coupled either to a small BEC or to a single atom.
For an atom cloud, three-body collisional loss is the dominating dissipative process, which restricts the choice of the trap frequency. For $N=500$ atoms in an elongated trap with aspect ratio $\omega_a/\omega_\parallel=25$ that is resonant with the CNT fundamental mode considered above with frequency $\omega_m/2\pi=20~$kHz, we obtain an atomic loss rate of $\gamma_a=2\pi \times 13~$Hz which we assume to be also the value of the atomic decoherence rate $\gamma_\mathrm{a,dec}$. For a quality factor of $Q=10^6$ and $T=10~$mK we have a decoherence rate $\gamma_\mathrm{m,dec}=k_BT/\hbar Q=2\pi\times 210~$Hz. This is to be compared to the single-phonon single-atom coupling rate, which amounts to $g_0=2\pi\epsilon\times 35~$Hz, and for an ensemble of atoms, the coupling is collectively enhanced according to $g_N = g_0\sqrt{N}$.
On the other hand, for a single atom in a magnetic trap, the only dissipative processes are collisions with the background gas, technical current noise induced heating, or Johnson noise induced spin-flip loss. These effects are small or can be controlled, and we assume an atomic decoherence rate $\gamma_\mathrm{a,dec}=2\pi \times 1~$Hz.
Here, the trap frequency can be made as large as possible, and we choose $\omega_a=2\pi \times250~$kHz in a magnetic microtrap. For a CNT with $l=4.25~\mu$m such that $\omega_m=\omega_a$ we find a coupling rate $g_0= 2\pi\epsilon\times 800~$Hz, and for the same quality factor as above we again obtain $\gamma_\mathrm{m,dec}=2\pi\times 210~$Hz.
Comparing the coherent coupling $g=\{g_0,g_N\}$ and dissipative rates for both cases we obtain
\begin{equation}
(g,\gamma_\mathrm{m,dec},\gamma_\mathrm{a,dec})=2\pi\times \left\{\begin{array}{rl} 
(\epsilon\times 780,~210,13)~\textrm{Hz} \quad &\textrm{collective}\\
(\epsilon\times 800,~210,~1)~\textrm{Hz} \quad &\textrm{single atom.}\\
\end{array}\right.
\end{equation}
This is sufficient to enter the strong coupling limit both for a single atom and collectively for an atom cloud with comparable rates. Furthermore, the strong coupling condition is met for a feasible coupling strength parameter down to $\epsilon\simeq 0.3$. However, cryogenic ground state cooling will not be feasible for the low resonance frequency considered here, and the mechanical mode will still be thermally populated with a mean phonon number $n_\mathrm{th}\sim10^3$. Additional cooling e.g.\ via radiation pressure \cite{Thompson08,Favero08,Favero09} in an optical cavity could be a way to initialize the oscillator in the ground state.

\section{Coupling mediated by an optical lattice in free space}\label{sec:LatticeCoupling}

\begin{figure}[tb!]
\centering
\includegraphics[width=0.65\textwidth]{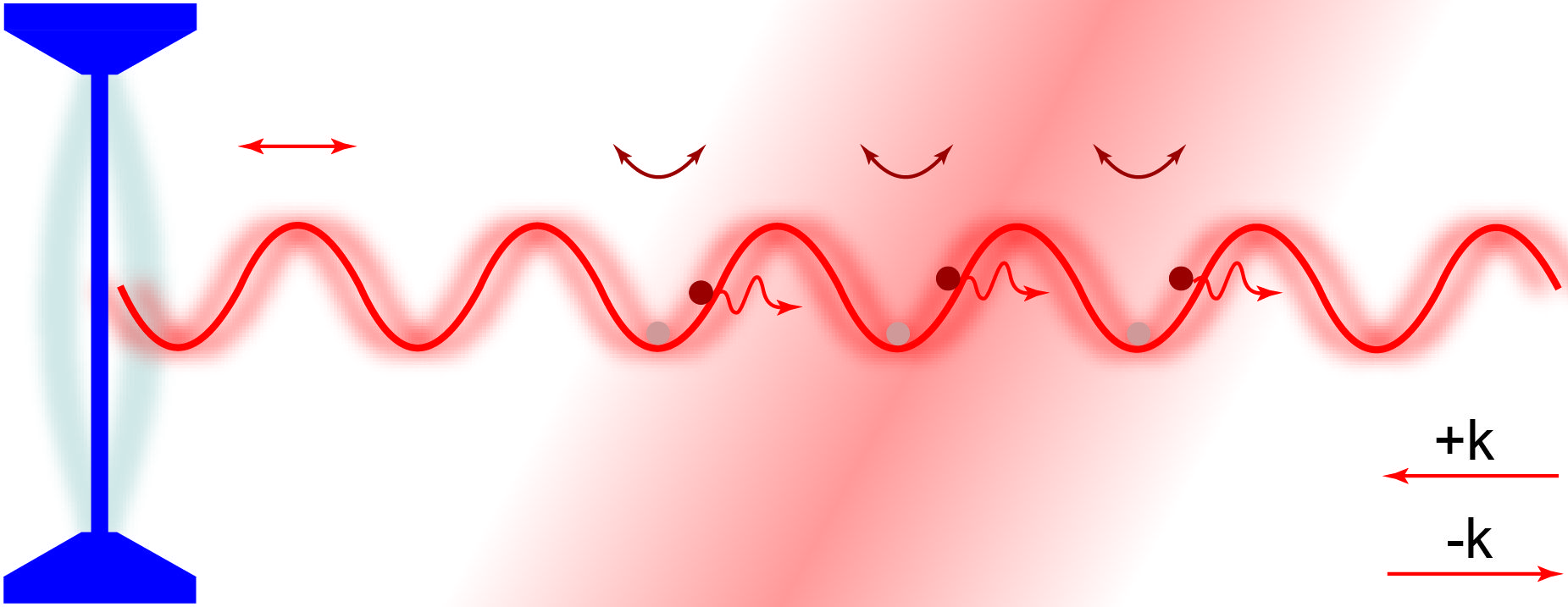}
\caption{\label{fig:LatticeCoupSchem} Schematic coupling mechanism: A lattice laser is retro reflected on a membrane oscillator to form a standing wave dipole trap. Oscillations of the membrane shake the lattice and excite c.o.m.\ motion of atoms trapped in the lattice. Laser cooling of the atoms provides sympathetic cooling of the oscillator.}
\end{figure}

Coupling many atoms collectively to a mechanical oscillator is another way to achieve strong coupling. In this section we discuss a scheme where an optical lattice provides a long-distance coupling between a micromechanical membrane oscillator and $N$ laser cooled atoms in the lattice \cite{Hammerer10}.
Consider a laser beam with wave vector $k$ that is retro-reflected on a membrane oscillator to form a 1D standing-wave optical potential of the form $U(z_a)=U_0\sin(k(z_a-z_m))^2$ with the potential depth $U_0$, suitable to trap laser cooled atoms. Movements of the membrane will move the standing wave, such that the membrane oscillations are directly transferred to an oscillation of the trapping potential of the atoms. This is a direct mechanical coupling with a coupling parameter $\epsilon=1$.
If the membrane oscillation frequency is resonant with the trap frequency of the atoms along the lattice direction, resonant excitation of the collective c.o.m.\ mode of the atoms leads to an energy transfer from the membrane to the atom cloud. Figure \ref{fig:LatticeCoupSchem} depicts the situation.

On the other hand, there is also a back-action of the atoms onto the membrane: The restoring force $F=-m\omega_a^2\delta z_a$ that attracts an atom displaced by $\delta z_a$ towards an anti-node of the standing wave arises from a coherent redistribution of photons between the two $k$-vectors components that make up the lattice \cite{Raithel98}.
For an oscillating atom, this leads to a modulation of the power of the beam impinging on the membrane
\begin{equation}
\Delta P = \frac{Fc}{2},
\end{equation} 
which results in a modulation of the radiation pressure force 
\begin{equation}
F_\mathrm{rp}=-\frac{2R\Delta P}{c}=-RF
\end{equation}
experienced by the membrane due to the reflection of the light with power reflectivity $R$. If the membrane is oscillating, it excites the atoms to collective motion through the coupling, and there is a fixed phase relation between atomic and membrane motion. The resulting radiation pressure modulation at the membrane will have the right phase lag to damp the membrane oscillation.
Furthermore, one can apply laser cooling to the trapped atoms and thereby continuously extract their gained energy. It was found in \cite{Hammerer10} that the resulting cooling rate of the membrane is given by
\begin{equation}
\Gamma_m=\gamma_m+\frac{4RNg_{0}^2}{\gamma_a^\mathrm{cool}}
\end{equation}
with the mechanical energy damping rate $\gamma_m=\omega_m/Q$, the atomic cooling rate $\gamma_a^\mathrm{cool}$, and the coupling rate $g_0 = \omega_m/2\sqrt{m/M}$. This leads to a steady state occupation of the membrane mode of
\begin{equation}
\bar{n}_{ss}=\frac{\gamma_m}{\Gamma_m}n_\mathrm{th}+\left(\frac{\gamma_a^\mathrm{cool}}{4\omega_m}\right)^2,
\end{equation}
where $n_\mathrm{th}\simeq k_B T /\hbar \omega_m$ is the thermal occupation of the membrane.
The motion of the membrane can in this way be sympathetically cooled via laser cooling of the atoms.

This scheme has several remarkable advantages.
Most prominently, the lattice provides a coupling that can bridge large distances (in principle hundreds of meters), limited only by retardation of the light field and by the frequency stability of the laser. This relaxes the technical effort considerably, as the mechanical oscillator does not have to be integrated into an ultracold atom vacuum setup. It also simplifies the implementation of a cryogenic environment for the oscillator.
On the atomic side, it is sufficient to superimpose laser cooling and an optical lattice, and no positioning is necessary. This simplifies and shortens the experimental cycle considerably.
Furthermore, laser cooling of the trapped atoms can be switched on and off, thereby giving access to both sympathetic cooling and the regime of coherent interaction.
Finally, the coupling strength can be collectively enhanced and thus made large by loading a large ensemble of atoms into the lattice. 

As an example, for a SiN membrane of dimensions $(l,w,h)=(150, 150, 0.05)~\mu$m with resonance frequency $\omega_m=2\pi\times 0.9~$MHz and effective mass $M=8\times 10^{-10}$~g one can expect $Q=10^7$ at $T=2~$K \cite{Thompson08} and a power reflectivity $R=0.3$ for $\lambda=780~$nm. For an ensemble of $N=3\times 10^8$ atoms \cite{Kerman00} one obtains a collectively enhanced coherent coupling rate $g_{N}=\sqrt{N}g_0=2\pi\times 3~$kHz, larger than all dissipative rates. Furthermore, if Raman sideband cooling is applied to the atoms, cooling factors of the order of $10^4$ can be achieved, sufficient for ground state cooling of a membrane precooled to below $500~$mK.

Already the simple picture given here shows the remarkable aspect of an asymmetric coupling: While the atoms experience a force $F$ that leads to a coupling constant $g_N=(\omega_m/2)\sqrt{Nm/M}$ as derived in Eq.~(\ref{eq:Coupling}), the membrane feels a force $RF$ which is reduced by the power reflectivity $R$, and thus also the coupling constant is reduced to $g_{m}=Rg_{N}$. A rigorous treatment of this effect as well as several dissipative processes such as radiation pressure noise call for an elaborate quantum description of the system. Such a description is given in \cite{Hammerer10} and it shows that the coupled atom-oscillator system considered here is an example of a cascaded quantum system.

\section{Coupling mediated by a high-finesse optical cavity}\label{sec:CavityCoupling}
\begin{figure}
\centering
\includegraphics[width=0.58\textwidth]{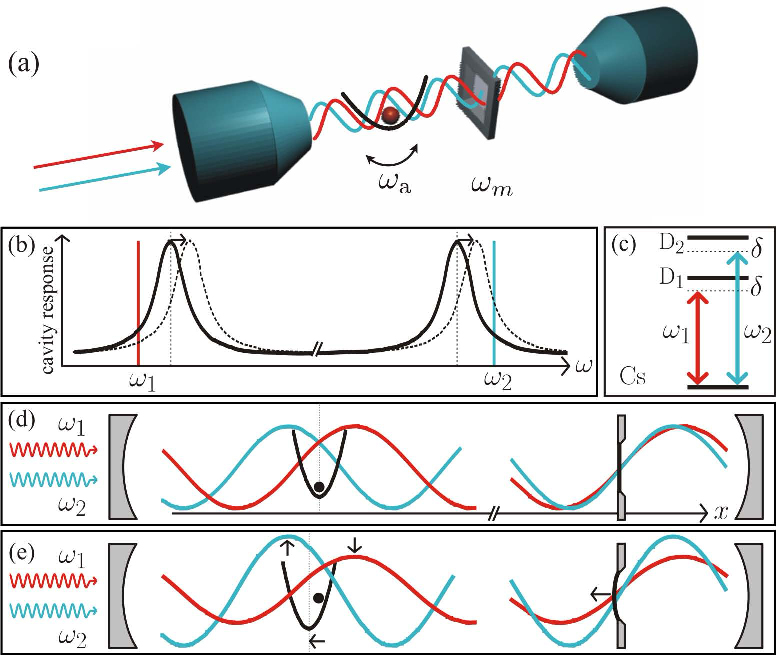} 
\caption{\label{fig:CavityCouplingSchem} (a) Coupling a single atom and a membrane via an optical cavity. (b) Cavity transmission spectrum and frequencies of two detuned lasers that mediate the coupling. (c) Relevant level scheme of Cs with the two laser frequencies, both red detuned from the atomic transitions to provide attractive dipole potentials. (d) The atom and the membrane are placed at specific positions within the spatial mode structure of the two standing wave cavity fields. (e) Small displacements of the membrane shift the cavity resonances and thus the relative intensity of the two fields (see (b)), thereby shifting the trap potential of the atom. Figure adapted from \cite{Hammerer09b}.
}
\end{figure}

The use of a high-finesse optical cavity to mediate the coupling represents an approach that does not require collective enhancement to achieve strong coupling. In \cite{Hammerer09b}, a scheme was elaborated that achieves a coupling that does not depend on the mass ratio $\sqrt{m/M}$, thereby enabling strong coupling between a single atom and a micromechanical membrane oscillator. The system combines an optomechanical setup with a membrane inside a cavity \cite{Thompson08} and a cavity QED system with a single trapped Cs atom \cite{Miller05}, Fig. \ref{fig:CavityCouplingSchem} (a) shows the setting.

The cavity light field interacts with the membrane via radiation pressure while it serves as a standing wave dipole trap for the atom. On the other hand, atom and membrane act as dispersive media that shift the cavity resonance depending on their position in the cavity light field. 
To realize an interaction that is linear in the amplitudes of the atom and the membrane, two different laser fields ($\omega_1,\omega_2$) are required with opposite detunings from the cavity resonance (see Fig.~\ref{fig:CavityCouplingSchem} b), but both red detuned from atomic transitions to provide an attractive potential for the atom (see Fig.~\ref{fig:CavityCouplingSchem} (c)). In this configuration, the motion of the atom or the membrane leads to a change in intracavity intensity, proportional to the gradient of the intensity at their respective positions. To achieve the desired coupling, the membrane has to be placed at a point of maximal intensity gradient with similar slope for both cavity fields, while the atom is trapped at a position where both fields have close to maximum but opposite slopes of the intensity (see Fig.~\ref{fig:CavityCouplingSchem} (d)). If the membrane is displaced from its equilibrium position, the blue detuned laser comes closer to resonance and generates a deeper lattice potential, while the red detuned laser becomes further detuned and thus weaker. This gives rise to a shift of the atomic trapping potential (Fig.~\ref{fig:CavityCouplingSchem} (e)) and results in the desired coupling of the type $H_\mathrm{int}=\hbar g_0 (\hat{a}^{\dagger}\hat{b}+\hat{a}\hat{b}^{\dagger})$.

A feasible set of parameters including a membrane of dimensions $(100\times100\times0.05)~\mu$m with $\omega_m=2\pi\times 1.3~$MHz and $Q=10^7$ at $T=2~$K, coupled to a single Cs atom through a cavity of finesse $F=2\times 10^5$, achieves a coupling rate $g_0=2\pi\times 45~$kHz and decoherence rates a factor 10 smaller, firmly entering the strong coupling regime.

This scheme achieves by far the largest coupling rate, even though the mechanical oscillator is rather large and the mass of the atom differs by 12 orders of magnitude. Since the required parameters have already been achieved in individual experiments \cite{Thompson08,Miller05}, the scheme appears as a promising, yet technologically challenging way to achieve strong coupling between a single atom and a mechanical oscillator at the quantum level.

\section{Magnetic coupling to the atomic spin}\label{sec:MagneticCoupling}
Coupling to internal states of the atoms has the advantage that high-frequency mechanical oscillators can be employed, as the oscillator vibrations now should be resonant e.g.\ with Zeeman transitions that can have frequencies in the MHz to GHz range. Additionally, the internal levels can be chosen and modified (e.g.\ by state-selective microwave dressing \cite{Boehi09}) such that the atoms act as effective two-level systems. This is important for the generation and readout of quantum states, because in a two-level system, single excitation quanta and coherent superposition states can be prepared by driving Rabi oscillations with classical fields. Furthermore, the quantum control of internal states is more advanced than for motional states (at least for neutral atoms), and one can benefit from the exceptionally long coherence lifetimes of internal states.

We consider the scenario of a nanomechanical cantilever with a nanomagnet on the tip, whose magnetic field couples to the spin of a trapped atom or BEC and drives Zeeman transitions. This coupling has been studied both theoretically \cite{Treutlein07,Geraci09,Singh10,Joshi10} and in a first experiment involving atoms in a vapor cell \cite{Wang06} (see below). Figure \ref{fig:MagCoupSchem} shows the situation considered in \cite{Treutlein07} with the transitions of $\Rb$ atoms involved.
The magnetic field $\mathbf{B}_r$ of the nanomagnet transduces mechanical oscillations of the cantilever into magnetic field oscillations at the position of the atoms. The amplitude of the field oscillations is linked to the cantilever oscillation $b(t)$ via the field gradient $G_m$,
\begin{equation}
\mathbf{B}_r(t) = b(t)G_m \mathbf{e}_x.
\end{equation}
The interaction of the atomic spin with $\mathbf{B}_r(t)$ is described by the Zeeman Hamiltonian
\begin{equation}\label{eq:HZ}
H_Z=-\boldsymbol{\mu} \cdot \mathbf{B}_r(t)=\mu_B g_F F_x G_m b(t),
\end{equation}
where $\boldsymbol{\mu}=-\mu_Bg_F \mathbf{F}$ is the magnetic moment and $\mathbf{F}$ the atomic spin.
The field direction of $\mathbf{B}_r$ is chosen orthogonal to the field in the trap center $\mathbf{B}_0$ set by an external source, such that the oscillations can drive spin-flip transitions between neighboring Zeeman sublevels (see Fig.~\ref{fig:MagCoupSchem} b). The spin-flip transition rate becomes sizable when the field oscillations at frequency $\omega_m$ are resonant with the Larmor frequency $\omega_L = \mu_B |g_F|B_0/\hbar$ of the atoms in the trap, set by the value of $B_0$ in the trap center. Since $B_0$ can be easily varied experimentally, the resonance condition $\omega_m=\omega_L$ can be fulfilled for a large range of cantilever frequencies. Furthermore, by rapidly changing $B_0$, the coupling can be switched on and off.

\begin{figure}[t]
\centering
\includegraphics[width=0.9\textwidth]{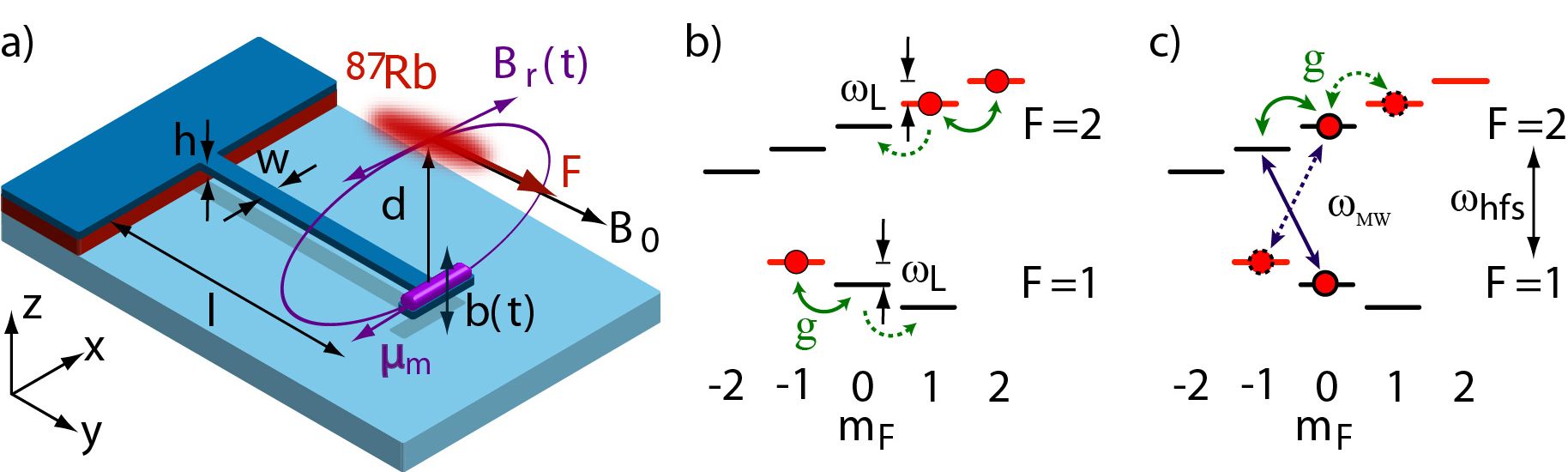}
\caption{\label{fig:MagCoupSchem} (a) Schematic of the magnetic coupling of a nanomechanical resonator to a single atom or a BEC. A nanomagnet on the tip of the cantilever transduces mechanical motion into magnetic field oscillations. Figure adapted from \cite{Treutlein07}. 
 (b) Ground state levels of $^{87}$Rb, magnetically trappable states are indicated in red. Magnetic field oscillations at the Larmor frequency $\omega_L$ lead to a coupling $g$ between different Zeeman sublevels. (c) Coupling via two-photon transitions with an additional microwave field permits to connect clock states with identical (dashed) or vanishing (solid) magnetic moment. }
\end{figure}


The magnet can be either a single domain ferromagnet with the magnetic moment $\boldsymbol{\mu}_m$ spontaneously oriented along its long axis due to the shape anisotropy, or a grain of a permanent magnet.
Approximating the magnet by a magnetic dipole, we have $G_m = 3\mu_0 |\boldsymbol{\mu}_m| / 4 \pi d^4$. By changing $d$ and the magnet dimensions, $G_m$ can be adjusted. Equation~(\ref{eq:HZ}) suggests that the strength of the atom-nanoresonator coupling can be maximized by increasing $G_m$ as much as possible. However, the atoms experience a force in this field gradient, and an excessively large $G_m$ would strongly distort the trapping potential, leading to a similar limitation as discussed for the direct mechanical coupling in Sec.~\ref{sec:DirectCoupling}. To mitigate distortion, two compensation magnets can be placed next to the coupling magnet with identical direction of magnetization. This reduces the static field gradient at the location of the atoms, while the oscillatory field $\mathbf{B}_r(t)$ remains unaffected as the compensation magnets do not oscillate.

It was already demonstrated \cite{Hofheinz09} that preparation of arbitrary quantum states in a harmonic oscillator can be achieved by a switchable, linear coupling to a two-level system with the Law-Eberly protocol \cite{Law96}. In the present scheme, a single atom is an effective two-level system if other transitions are sufficiently detuned. This can be achieved intrinsically via the quadratic Zeeman effect, or controllably via state-selective microwave dressing \cite{Boehi09}.
Transitions between two states $|g\rangle$ and  $|e\rangle$ of the atom can be described by creation and annihilation operators $\hat{\sigma}^+=|e\rangle\langle g|$ and $\hat{\sigma}^-=|g\rangle\langle e|$, and with the linear coupling to the mechanical oscillator, the system is analog to the well-known Jaynes-Cummings model described by the Hamiltonian \cite{Shore93}
\begin{equation}\label{eq:HJC}
H = \hbar \omega_m  \hat{b}^{\dagger}  \hat{b}  + \hbar \omega_L  \hat{\sigma}^+\hat{\sigma}^- + \hbar g_0 ( \hat{\sigma}^+  \hat{b}  +  \hat{\sigma}^- \hat{b}^{\dagger}),
\end{equation}
where $g_0=\mu_B G_m b_\mathrm{qm}/\sqrt{8} \hbar$
is the single-atom single-phonon coupling constant for the state pair $|e\rangle=|1,-1\rangle \rightarrow |g\rangle=|1,0\rangle$.
In the field of cavity QED, Eq.~(\ref{eq:HJC}) usually describes the coupling of an atom to the electromagnetic field of a single mode of an optical or microwave cavity \cite{Shore93}. Here, it describes the coupling to the phonon field of a single mode of mechanical oscillations. In this sense, this is a mechanical cavity QED system. In the case of an ensemble of atoms coupling collectively to the oscillator, the description is given by the Tavis-Cummings model \cite{Treutlein07,Shore93}.

This system can achieve the strong coupling regime both for a single atom and for a small cloud of atoms. A single $\Rb$ atom trapped at a distance $d=250~$nm from a cantilever with $(l,w,h)=(8,0.3,0.05)~\mu$m and $\omega_m=2\pi\times 2.8~$MHz carrying a single domain Co nanomagnet of dimensions $(250,50,80)~$nm achieves a coupling rate $g_0=2\pi\times 60~$Hz.
The expected decoherence rate of the atoms depends strongly on the involved magnetic levels.
For coherence times on the order of seconds, magnetic sublevels with identical ($|1,-1\rangle \rightarrow |2,1\rangle$) or vanishing ($|1,0\rangle \rightarrow |2,0\rangle$) magnetic moment  have to be used \cite{Harber02}, see Fig.~\ref{fig:MagCoupSchem} c. The latter state pair also avoids trap distortion due to the coupling magnet, but it requires an optical or electrodynamical trap to hold the atom \cite{Treutlein07,Katori04}. To connect states in different hyperfine manifolds, a two-photon process is required with an additional strong microwave field at $\omega_\mathrm{MW}\simeq \omega_\mathrm{hfs}$. If the parameters are adjusted properly, the two-photon coupling reduces the effective coupling rate only by a factor of $\approx 3$. 
A sufficiently low decoherence rate of the cantilever to achieve the strong coupling condition $g_0>\gamma_\mathrm{a,dec},\gamma_\mathrm{m,dec}$ again requires a low bath temperature $T=10~$mK and a very high quality factor $Q=10^7$. Similar to the previous discussion for the nanotube in Sec.~\ref{sec:CNTCoupling}, cryogenic ground state cooling of the cantilever is not available. Further cooling could be achieved by a state swap with a BEC of $N$ atoms collectively coupled to the oscillator.
Excitations of the oscillator can be coherently transferred to the atoms prepared in state $|g\rangle^{\otimes N}$, such that after an interaction time $t=\pi/2g_0\sqrt{N}$ the oscillator is in a state with reduced mean phonon number.
Alternatively, by making use of atomic transitions between the two hyperfine manifolds $F=1$ and $F=2$ with a transition frequency $\omega_\mathrm{hfs}/2\pi\simeq 6.8~$GHz, one could employ GHz mechanical oscillators, for which cryogenic ground state cooling can be achieved \cite{OConnell10}.

Based on the magnetic coupling mechanism described above, a scalable experimental setup was proposed in \cite{Geraci09}, where the atoms are trapped in an optical lattice and each lattice site couples to an individual mechanical oscillator. This could be used for atomic state manipulation with single lattice site resolution, for the operation of quantum gates, or for the entanglement of distant mechanical oscillators \cite{Geraci09,Joshi10}.

A magnetic interaction was also used in the first experiment that demonstrated a coupling between a mechanical oscillator and an atomic system \cite{Wang06}. There, a microfabricated cantilever with a magnetic tip was excited to coherent motion and coupled to a gas of Rb atoms in a vapor cell. The oscillating magnetic field of the tip causes spin precession of the atoms, which can be detected as an intensity modulation of a detection laser. This scheme can serve for magnetic field sensing, and a sensitivity of $10~$nT$/\sqrt{\mathrm{Hz}}$ was demonstrated.

Finally, magnetic interactions can also be used to couple a mechanical oscillator to a solid-state ``artificial atom'' such as a nitrogen-vacancy center (NV) in diamond \cite{Rabl09,Rabl10}. The advantage compared to ultracold trapped atoms is that a large coupling strength $g_0/2\pi\simeq100~$kHz can be achieved because the distance between the magnetic tip and the NV center can be made very small  ($d=25~$nm) and distortion of trapping potentials is not an issue. To compete with the typically faster decoherence of the NV ground states, level dressing was proposed to tune the coupled system to a sweet spot with strongly suppressed magnetic field sensitivity and thus reduced decoherence rate, sufficient to achieve the strong coupling regime \cite{Rabl09}.

\section{Conclusion}\label{sec:Conclusion}
Ultracold atoms coupled to mechanical oscillators represent a novel type of hybrid mechanical system that can benefit from the exceptional coherence properties and the versatile techniques for quantum manipulation of trapped atoms. This could enable readout, cooling, and coherent manipulation of mechanical systems via the well-developed tools of atomic physics. A multifaceted set of theoretical proposals shows that several different approaches have the potential to achieve strong coherent coupling between atoms and micro- or nanostructured oscillators. 
The central challenge is that the overall dynamics of the system is relatively slow, so that exceptional coherence properties of the mechanical oscillator are required. 
However, in light of the recent experimental progress with high-quality mechanical oscillators at low temperatures it seems feasible that this challenge can be met. 

Experimental work on coupled atom-oscillator systems is still at an early stage, but first experiments have already demonstrated different coupling mechanisms  \cite{Wang06,Hunger10a}. Moreover, some of the proposed schemes can be implemented with moderate technical effort  \cite{Hammerer10} and enable the observation of backaction of the atoms onto the mechanical oscillator already in room temperature setups. 
Coupling schemes based on high-finesse optical cavities promise access to a regime of strong coherent coupling with state-of-the-art experimental parameters \cite{Hammerer09b,Wallquist10}, and thus offer a realistic perspective for experiments on the quantum physics of massive mechanical objects.  

\subsection*{Acknowledgements}
We acknowledge fruitful collaborations and discussions with J. Reichel, J. P. Kotthaus, D. K{\"o}nig, K. Hammerer, M. Wallquist, C. Genes, K. Stannigel, P. Zoller, M. Ludwig, F. Marquardt, H. J. Kimble, J. Ye, P. Rabl, M. Lukin, and D. Leibfried. This work was supported by the excellence cluster Nanosystems Initiative Munich (NIM) and by the European Community's Seventh Framework Programme through the Integrating Project AQUTE.


\end{document}